\documentclass[prd,twocolumn,amsmath,amssymb,superscriptaddress]{revtex4-2}
\usepackage{graphicx}
\usepackage{hyperref}
\usepackage[nodisplayskipstretch]{setspace}

\def\tac{\langle \sigma_\text{ann} v\rangle}
\def\fermi{\textit{Fermi}-LAT}
\def\deg{^{\circ}}

\def\mdm{m_\text{DM}}

\def\UCLA{\affiliation{Department of Physics and Astronomy, University of California, Los Angeles, CA 90095, USA} \email[Contact author: ]{rene@astro.ucla.edu}}

\makeatletter
\newcommand{\hathat}[1]{\begingroup
  \let\macc@kerna\z@ \let\macc@kernb\z@ \let\macc@nucleus\@empty%
  \hat{\raisebox{.35ex}{\vphantom{\ensuremath{#1}}}\smash{\hat{#1}}}%
\endgroup} \makeatother

\begin{document}

\title{VERITAS Limits on Dark Matter Annihilation in the Galactic Center Halo}

\author{A.~Archer}\affiliation{Department of Physics and Astronomy, DePauw University, Greencastle, IN 46135-0037, USA}
\author{P.~Bangale}\affiliation{Department of Physics, Temple University, Philadelphia, PA 19122, USA}
\author{J.~T.~Bartkoske}\affiliation{Department of Physics and Astronomy, University of Utah, Salt Lake City, UT 84112, USA}
\author{W.~Benbow}\affiliation{Center for Astrophysics $|$ Harvard \& Smithsonian, Cambridge, MA 02138, USA}
\author{N.~R.~Bond}\affiliation{School of Physics, University College Dublin, Belfield, Dublin 4, Ireland}
\author{J.~H.~Buckley}\affiliation{Department of Physics, Washington University, St. Louis, MO 63130, USA}
\author{Y.~Chen}\UCLA
\author{J.~L.~Christiansen}\affiliation{Physics Department, California Polytechnic State University, San Luis Obispo, CA 94307, USA}
\author{A.~J.~Chromey}\affiliation{Center for Astrophysics $|$ Harvard \& Smithsonian, Cambridge, MA 02138, USA}
\author{A.~Duerr}\affiliation{Department of Physics and Astronomy, University of Utah, Salt Lake City, UT 84112, USA}
\author{M.~Errando}\affiliation{Department of Physics, Washington University, St. Louis, MO 63130, USA}
\author{M.~Escobar~Godoy}\affiliation{Santa Cruz Institute for Particle Physics and Department of Physics, University of California, Santa Cruz, CA 95064, USA}
\author{J.~Escudero~Pedrosa}\affiliation{Center for Astrophysics $|$ Harvard \& Smithsonian, Cambridge, MA 02138, USA}
\author{S.~Feldman}\UCLA
\author{Q.~Feng}\affiliation{Department of Physics and Astronomy, University of Utah, Salt Lake City, UT 84112, USA}
\author{S.~Filbert}\affiliation{Department of Physics and Astronomy, University of Utah, Salt Lake City, UT 84112, USA}
\author{L.~Fortson}\affiliation{School of Physics and Astronomy, University of Minnesota, Minneapolis, MN 55455, USA}
\author{A.~Furniss}\affiliation{Santa Cruz Institute for Particle Physics and Department of Physics, University of California, Santa Cruz, CA 95064, USA}
\author{W.~Hanlon}\affiliation{Center for Astrophysics $|$ Harvard \& Smithsonian, Cambridge, MA 02138, USA}
\author{O.~Hervet}\affiliation{Santa Cruz Institute for Particle Physics and Department of Physics, University of California, Santa Cruz, CA 95064, USA}
\author{C.~E.~Hinrichs}\affiliation{Center for Astrophysics $|$ Harvard \& Smithsonian, Cambridge, MA 02138, USA and Department of Physics and Astronomy, Dartmouth College, 6127 Wilder Laboratory, Hanover, NH 03755 USA}
\author{J.~Holder}\affiliation{Department of Physics and Astronomy and the Bartol Research Institute, University of Delaware, Newark, DE 19716, USA}
\author{D.~Huang}\affiliation{Department of Physics and Astronomy and the Bartol Research Institute, University of Delaware, Newark, DE 19716, USA}
\author{T.~B.~Humensky}\affiliation{Department of Physics, University of Maryland, College Park, MD, USA and NASA GSFC, Greenbelt, MD 20771, USA}
\author{M.~Iskakova}\affiliation{Department of Physics, Washington University, St. Louis, MO 63130, USA}
\author{W.~Jin}\UCLA
\author{E.~Joshi}\affiliation{DESY, Platanenallee 6, 15738 Zeuthen, Germany}
\author{M.~Kertzman}\affiliation{Department of Physics and Astronomy, DePauw University, Greencastle, IN 46135-0037, USA}
\author{M.~Kherlakian}\affiliation{Fakult\"at f\"ur Physik \& Astronomie, Ruhr-Universit\"at Bochum, D-44780 Bochum, Germany}
\author{D.~Kieda}\affiliation{Department of Physics and Astronomy, University of Utah, Salt Lake City, UT 84112, USA}
\author{T.~K.~Kleiner}\affiliation{DESY, Platanenallee 6, 15738 Zeuthen, Germany}
\author{N.~Korzoun}\affiliation{Department of Physics and Astronomy and the Bartol Research Institute, University of Delaware, Newark, DE 19716, USA}
\author{S.~Kundu}\affiliation{Department of Physics and Astronomy, University of Alabama, Tuscaloosa, AL 35487, USA}
\author{M.~J.~Lang}\affiliation{School of Natural Sciences, University of Galway, University Road, Galway, H91 TK33, Ireland}
\author{M.~Lundy}\affiliation{Columbia Astrophysics Laboratory, Columbia University, New York, NY 10027, USA}
\author{G.~Maier}\affiliation{DESY, Platanenallee 6, 15738 Zeuthen, Germany}
\author{M.~R.~Martin}\affiliation{Department of Physics and Astronomy, University of Utah, Salt Lake City, UT 84112, USA}
\author{P.~Moriarty}\affiliation{School of Natural Sciences, University of Galway, University Road, Galway, H91 TK33, Ireland}
\author{R.~Mukherjee}\affiliation{Department of Physics and Astronomy, Barnard College, Columbia University, NY 10027, USA}
\author{W.~Ning}\UCLA
\author{M.~Ohishi}\affiliation{Institute for Cosmic Ray Research, University of Tokyo, 5-1-5, Kashiwa-no-ha, Kashiwa, Chiba 277-8582, Japan}

\author{R.~A.~Ong}
\UCLA

\author{A.~Pandey}\affiliation{Department of Physics and Astronomy, University of Utah, Salt Lake City, UT 84112, USA}
\author{M.~Pohl}\affiliation{Institute of Physics and Astronomy, University of Potsdam, 14476 Potsdam-Golm, Germany and DESY, Platanenallee 6, 15738 Zeuthen, Germany}
\author{J.~Quinn}\affiliation{School of Physics, University College Dublin, Belfield, Dublin 4, Ireland}
\author{P.~L.~Rabinowitz}\affiliation{Department of Physics, Washington University, St. Louis, MO 63130, USA}
\author{K.~Ragan}\affiliation{Physics Department, McGill University, Montreal, QC H3A 2T8, Canada}
\author{P.~T.~Reynolds}\affiliation{Department of Physical Sciences, Munster Technological University, Bishopstown, Cork, T12 P928, Ireland}
\author{E.~Roache}\affiliation{Center for Astrophysics $|$ Harvard \& Smithsonian, Cambridge, MA 02138, USA}

\author{J.~L.~Ryan}
\affiliation{SLAC National Accelerator Laboratory/Kavli Institute for Particle Astrophysics and Cosmology, Menlo Park, CA 94025, USA}
\email[Contact author: ]{jlryan@slac.stanford.edu}

\author{I.~Sadeh}\affiliation{DESY, Platanenallee 6, 15738 Zeuthen, Germany}
\author{L.~Saha}\affiliation{Center for Astrophysics $|$ Harvard \& Smithsonian, Cambridge, MA 02138, USA}
\author{H.~Salzmann}\affiliation{Santa Cruz Institute for Particle Physics and Department of Physics, University of California, Santa Cruz, CA 95064, USA}
\author{M.~Santander}\affiliation{Department of Physics and Astronomy, University of Alabama, Tuscaloosa, AL 35487, USA}
\author{G.~H.~Sembroski}\affiliation{Department of Physics and Astronomy, Purdue University, West Lafayette, IN 47907, USA}
\author{D.~Tak}\affiliation{Department of Physics, Kyung Hee University, Seoul 02447, Republic of Korea}
\author{J.~V.~Tucci}\affiliation{Department of Physics, Indiana University Indianapolis, Indianapolis, Indiana 46202, USA}
\author{V.~V.~Vassiliev}\UCLA
\author{D.~A.~Williams}\affiliation{Santa Cruz Institute for Particle Physics and Department of Physics, University of California, Santa Cruz, CA 95064, USA}
\author{S.~L.~Wong}\affiliation{Physics Department, McGill University, Montreal, QC H3A 2T8, Canada}
\author{T.~Yoshikoshi}\affiliation{Institute for Cosmic Ray Research, University of Tokyo, 5-1-5, Kashiwa-no-ha, Kashiwa, Chiba 277-8582, Japan}


\collaboration{VERITAS Collaboration}

\date{\today}

\begin{abstract}
Well-motivated models of particle dark matter predict its self-annihilation into Standard Model end states, which may be detected by particle detectors or telescopes.
We conduct an indirect dark matter search using gamma rays detected by the VERITAS telescope from within 2$\deg$ of the Galactic Center.
A template-based approach is used for background estimation, increasing our sensitivity to sources of gamma-ray emission with spatial extents comparable to the 3.5$\deg$ VERITAS field of view.
The analysis benefits from data taken exclusively at large zenith angles, which results in an increased effective area at energies above $\sim$1 TeV, compared to small zenith angle observations.
No significant gamma-ray signal is observed in the search region, resulting in upper limits on the end-state-dependent dark matter annihilation cross sections.
Cross sections as low as $\tac = 1.77\times 10^{-25} \text{ cm}^3 \text{ s}^{-1}$ are excluded at 95\% confidence assuming complete annihilation into $\tau^+ \tau^-$, an Einasto profile, and a dark matter mass of 1 TeV.
\end{abstract}


\maketitle

\section{Introduction} \label{sec:intro}
Our cosmological paradigm includes a dark matter component constituting approximately 84\% of the Universe's matter density \cite{planck20}.
Dark matter, in addition to being a necessary ingredient in most cosmological models, simultaneously resolves numerous other astrophysical anomalies.
However, dark matter has to date eluded explanation from any Standard Model particle or astrophysical source.
Its nature, particle or otherwise, remains unknown.

Theoretically-appealing candidates, 
such as weakly-interacting massive particles (WIMPs), predict the self-annihilation of dark matter into Standard Model end states---the observation of which would constitute an indirect detection of dark matter particles (e.g. Ref. \cite{jungman96}).
Among the most promising avenues of indirect detection are gamma-ray searches at the Galactic Center, where the dark matter density in our Galaxy is expected to peak \citep{bergstrom98}.
At GeV energies, such searches have already revealed the puzzling Galactic Center excess \citep{goodenough09,hooper11,ackermann17}, whose dark-matter interpretation is complicated by its morphology and astrophysical foregrounds (e.g. Refs. \cite{macias18,mcdermott23,song24}), while observations at TeV energies have placed the most stringent constraints on a variety of dark matter models \citep{hess18,hess22,abe23}.

Such dark matter searches must contend with the presence of a variety of astrophysical sources within a few degrees of the Galactic Center.
Even at TeV energies, sources of emission are believed to include supernova remnants, diffuse emission from cosmic-ray interactions with gas, and a source in the direction of the supermassive black hole designated Sagittarius A* (Sgr A*) \cite{adams21}.
Ground-based gamma-ray observations additionally face a background due to cosmic-ray interactions with the atmosphere.
These effects complicate the search for a gamma-ray signal from the Milky Way's dark matter halo, whose brightest region may be fairly spatially extended, in principle.
The present work develops a methodology that can handle these properties, by means of background templates.
This methodology is well-suited for the analysis of extended sources, even in the presence of other gamma-ray sources in the same field of view.

The data, experimental apparatus, and methodology are described in \S \ref{sec:datamethods}.
A newly-developed, template-based background estimation method is described, and its validity demonstrated, in \S \ref{sec:tbm}.
This method is then applied to gamma-ray observations of the Galactic Center in \S \ref{sec:analysis}.
Signal significance and dark matter annihilation cross sectional limits are evaluated in \S \ref{sec:res} and discussed in \S \ref{sec:disc}. Conclusions are drawn in \S \ref{sec:conc}.

\section{Data and Methodology} \label{sec:datamethods}

\subsection{Data} \label{sec:data}
Data were taken with the Very Energetic Radiation Imaging Telescope Array System (VERITAS) \cite{weekes02}, an imaging atmospheric Cherenkov telescope (IACT) array located at the Fred Lawrence Whipple Observatory in southern Arizona at $31\deg \; 40'$ N, $110\deg \; 57'$ W, approximately 1.3 km above sea level.
When observing at small zenith angles, VERITAS is sensitive to gamma rays with energies from 100 GeV to over 30 TeV, achieving an energy resolution of 15--25\% and an angular resolution $<0.1\deg$ at 1 TeV.
VERITAS can make a statistically-significant detection (greater than five standard deviations, or $5 \sigma$, above background)
of a source with 1\% the gamma-ray flux from the Crab nebula in $< 25$ hours of observing time.

Sgr A* was observed for 177.8 hours between 2010 April and 2022 June with VERITAS. 
An additional 50.8 hours of off-source observations were taken on a nearby patch of sky with no known gamma-ray sources and few bright stars, which we refer to as Sgr A* Off.
All observations were taken with a telescope pointing offset of $0.5\deg$ or $0.7\deg$ from the nominal source position, which is
(RA, Dec)=$(266.417\deg,-29.0078\deg)$ for Sgr A* and 
(RA, Dec)=($260.917\deg,-29.0078\deg)$ for Sgr A* Off, in equatorial coordinates (J2000).
As a consequence of VERITAS' location, the data are taken between late March and early July, at elevation angles $\lesssim 30\deg$.
Significant upgrades to the VERITAS cameras and trigger system were made in summer 2012 \citep{kieda13} -- our data taken before and after the upgrade are referred to as V5 and V6, respectively. The V5 and V6 datasets have substantially different instrumental response functions.
The VERITAS camera gains and mirror reflectivities also evolved over the observation period \citep{adams22}.
 
Data are required to have been taken in good weather and to have all four telescopes participating.
Data runs exhibiting large anomalies in triggering rates, sky temperatures, or other metrics are not included in the analysis. 
Runs showing short segments of anomalous behavior (e.g. due to clouds or car headlights) are used, with those segments excluded.
The total exposure times remaining after applying data quality selection criteria and deadtime corrections are 154.3 hours for Sgr A* and 44.0 hours for Sgr A* Off.

Several bright stars and gamma-ray sources are present in our observational fields of view.
In order to carry out a dark matter search, we exclude regions around these sources by ignoring events in these regions. 
In the Sgr A* Off field, we exclude $0.3\deg$ around the bright stars d Oph and HIP 85442.
In the Sgr A* field, we exclude $0.3\deg$ around the bright star X Sgr, $0.4\deg$ around gamma-ray sources Sgr A*, G0.9+0.1, and HESS J1746--285, and $0.6\deg$ around HESS J1744--303 \citep{hess18,adams21}.
The region $|b|<0.3\deg$, covering emission along the Galactic plane, is also excluded.

\subsection{Instrument Response Functions}
Data taken at large zenith angles (LZA; $\gtrsim 40\deg$) require instrument response functions (IRFs) different from those used for smaller zenith angles \citep{buchovecky19}.
The distance from the VERITAS telescopes to the gamma-ray-initiated electromagnetic particle cascade (``air shower'') points of origin observed at LZA are typically greater than when observing near zenith, resulting in both higher effective areas and a higher energy threshold \citep{sommers87}.
The standard VERITAS event reconstruction pipeline \citep{cogan08}, which uses stereoscopic information to geometrically reconstruct air shower direction, is also less accurate at LZA.
The energy threshold for our analysis is 2 TeV, as determined by the IRFs.
Effective areas reach $\sim$4$\times 10^5 \text{ m}^2$ above 10 TeV.


The IRFs used in the present analysis were used previously in the Galactic Center analysis of Ref. \cite{adams21}.
As described in Ref. \cite{buchovecky19},
the IRFs are implemented using boosted decision trees (BDTs)
trained on simulated air showers.
The BDTs estimate the event direction and energy using the Hillas image parameters length, width, and size \citep{hillas96}, as well as the arrival time gradient across the image and the integrated charge lost due to image truncation at the camera edges.
Separate BDTs are trained for different observation zenith angles, azimuth angles, and night sky background levels.
We obtain an angular resolution ${\sim}0.12\deg$ and an energy resolution ${\sim}19\%$ above 2 TeV for LZA observations using these specialized IRFs. 
Effective area lookup tables are calculated from the reconstructed, simulated events.
Gamma-hadron separation is accomplished with standard selection criteria (``cuts'') optimized for LZA observations of the Crab nebula---we note that these cuts are not necessarily optimal for the dark-matter-induced gamma-ray spectra and region of interest considered.
The IRFs do not account for the throughput calibration described in Ref. \cite{adams22}. Throughput corrections are instead applied later to the theoretical dark-matter-induced gamma-ray fluxes.

\subsection{Template Background Method} \label{sec:tbm}
IACT analyses are generically subject to a background of gamma-ray-like events that do not originate from gamma rays---mostly due to showers initiated by charged cosmic rays---that are not removed by gamma-hadron separation cuts.
Standard background-estimation methods, such as the ring or reflected-region methods (e.g. Ref. \cite{berge07}), require patches of sky in the same field of view (``off'' regions) that are large enough to contain a sufficient number of background events, without including many events from the gamma-ray source itself.
Spatially-extended gamma-ray sources covering a significant amount of the field of view may therefore preclude the use of such methods.

\begin{figure}[t]
    \centering
    \includegraphics[width=0.5\textwidth]{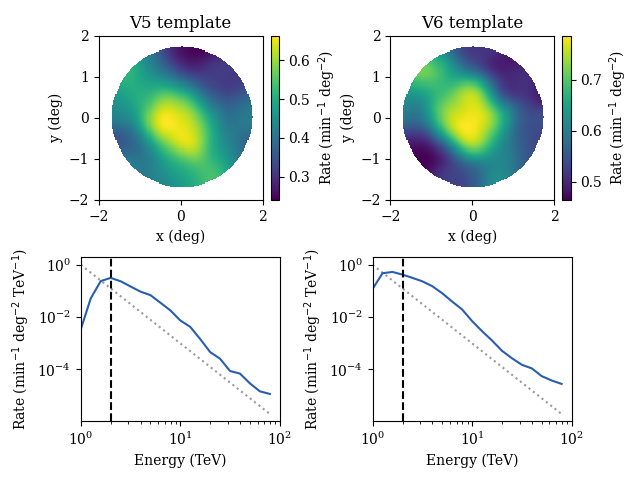}
    \caption{
Two-dimensional plots illustrating features of the background templates.
Shown are the event rates as a function of camera position for the V5 template (top left) and the V6 template (top right), integrated over energy.
Asymmetry can be observed along the y-axis, which aligns with zenith angle.
Background-rate energy spectra, averaged over spatial coordinates, are also shown for V5 (bottom left) and V6 (bottom right).
The V5 and V6 VERITAS epochs are described in Section \ref{sec:data}.
Gray dotted lines are power laws with index $-3$.
}
    \label{fig:templates}
\end{figure}

We have developed a background estimation method that does not use gamma-ray-like events in the field of view.
The method uses templates of gamma-ray-like events to estimate the expected number of events as a function of sky coordinates and energy, 
and so we refer to it as the template background method (TBM) \cite{ryan22}.
Similar approaches have been used in previous IACT analyses (e.g. Refs. \cite{mayer14,ziegler18,vovk18,kelleyhoskins19,knodlseder19,mohrmann19,acharyya21}).

The templates describe the spatial and spectral shapes of the background rate as a function of position and energy, with the overall normalization for each observation determined by observing conditions.
Templates are empirically derived from Sgr A* Off data, where the gamma-ray-like events are expected to be purely background-induced.
We use the same gamma-hadron separation procedure as Ref. \cite{adams21}.
Templates are stored in camera coordinates, covering a $3.4\deg$ diameter field of view divided into $0.025\deg$ angular bins, and 20 logarithmically-spaced energy bins between 1 TeV and 100 TeV.
To construct the templates, events from each Sgr A* Off data run are first accumulated in 3D histograms with the aforementioned binning. 
For energy bins with fewer than 5000 events, events from neighboring energy bins are added 
to reduce the effect of fluctuations due to low numbers of events---as a result, the higher-energy bins share an identical spatial template.
Live time is separately calculated for each spatial bin, accounting for exclusion regions, resulting in a 2D live-time map.
The arrays of event counts are smoothed in both spatial dimensions with a 2D Gaussian kernel with standard deviation $0.25\deg$---smaller values are found to produce templates that underestimate the number of events towards the edge of the field of view. 
The live-time map is convolved with the same Gaussian kernel to account for edge effects: the zero-valued region outside of the field of view causes values near the edges to be underestimated after smoothing.
Dividing, at each energy, the smoothed 2D arrays of counts by the smoothed live-time map produces a TBM template.
Such templates express rates 
that, for each observation, can be transformed into sky coordinates and multiplied by live time to yield the expected number of background counts at each position and energy.

Separate templates are derived for V5 and V6 data, due to anticipated differences between epochs, including differing energy thresholds, gamma/hadron separation cuts, and evolving instrument throughput factors.
Some template properties are illustrated in Figure \ref{fig:templates}.
A gradient is visible along the camera's $y$ axis, particularly in the V5 data, which aligns with the zenith axis and naturally enters background estimates.
The templates' differential energy spectra, obtained by integrating over both spatial dimensions, is approximately proportional to $E^{-3}$, where $E$ is energy, resembling the cosmic-ray spectrum.

\begin{figure}[t]
    \centering
    \includegraphics[width=0.5\textwidth]{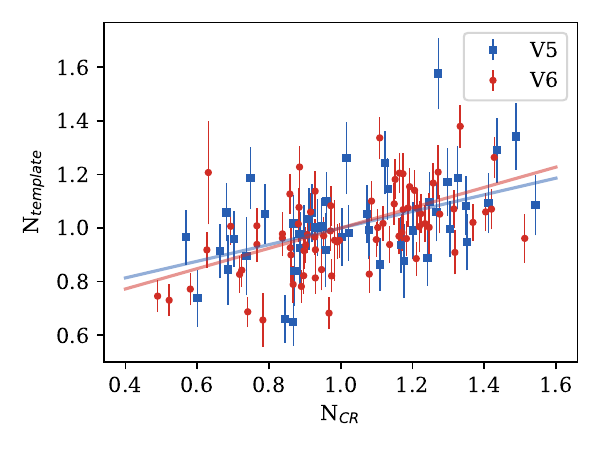}
    \caption{Template-scaling factor versus cosmic-ray template-scaling factor, for V5 and V6. Each data point corresponds to a single observation of Sgr A* Off. Straight lines represent linear fits of the form given in Eqs. \ref{eq:templatenorm} and \ref{eq:templatenorm2}. Error bars represent Gaussian approximations of the Poisson errors, though the fitting procedure uses the explicit Poisson probabilities.
}
    \label{fig:ncr}
\end{figure}

The variance in number of events of the individual Sgr A* Off observations exceeds the Poisson fluctuations expected for a constant background rate.
As is known from previous analyses that have used other background estimation techniques (e.g. \cite{hona21}), the background rate can vary significantly from run to run, depending on sky brightness and other observational parameters.
We observe correlations between the number of background counts and the number of cosmic-ray-like events, the zenith angle, and observation date.
We therefore introduce a template scaling factor $N_\text{template}$, calculated for each observation, that relates the expected background counts $n_\text{B}$ for a single observation to the template rate $T$ by
\begin{equation}
n_\text{B}(x,y,E) = N_\text{template}T(x,y,E)t_\text{live}
\end{equation}
where $x$ and $y$ are camera coordinates, $E$ is energy, and $t_\text{live}$ is live time.
For a background-only data run, the maximum-likelihood estimate of $N_\text{template}$ maximizes
\begin{equation} \label{eq:bglik}
L(N_\text{template}) = \prod_{i,j,k} \frac{n_\text{B}(x_i,y_j,E_k)^{n_{ijk}}}{n_{ijk}!}e^{-n_\text{B}(x_i,y_j,E_k)}
\end{equation}
where $n_{ijk}$ are the numbers of gamma-ray-like events contained in the bin corresponding to $(x_i,y_j,E_k)$.
To predict $N_\text{template}$ for a given run without using the gamma-ray-like events, we investigated a functional form for $N_\text{template}$ dependent on the number of cosmic-ray-like events---those excluded by gamma-hadron separation cuts---since the gamma-ray-like background is mainly cosmic-ray-induced.
The sum of cosmic-ray-like events cannot be used directly since it is affected by exclusion regions and non-uniform camera acceptance.
We instead use a cosmic-ray template scaling factor $N_\text{CR}$ using separate templates derived from the cosmic-ray-like events, calculated using a likelihood similar to Eq. \ref{eq:bglik} except with $n_{ijk}$ replaced with cosmic-ray-like events.
$N_\text{template}$ and $N_\text{CR}$ are calculated for each Sgr A* Off observation, and are plotted in Figure \ref{fig:ncr}.
Empirical relations for $N_\text{template}$ are derived separately for V5 and V6
\begin{align}
N_\text{template,V5} &= 1 + 0.310(N_\text{CR}-1) \label{eq:templatenorm} \\
N_\text{template,V6} &= 1 + 0.379(N_\text{CR}-1) \label{eq:templatenorm2}
\end{align}
by fitting a line to the template scaling factors derived from each run.
The relative uncertainty is found to be $\sim$13\% for both epochs, which determines the uncertainty on the background estimates.
This value is comparable to that of Ref. \cite{mohrmann19}, who determine normalizations for their three-dimensional background templates to within 12\%, using the parameters zenith angle, observation time, and their ``transparency coefficient.''
We explored the addition of terms with linear dependence on zenith angle and observation date, but found these to reduce fit residuals by $<1\%$, so did not include them.

\subsubsection{Template Background Method Statistics}
If a number of events in an on-source region $n_\text{on}$ are observed,
and the mean number of background events is Gaussian-distributed with mean $\mu_\text{B}$ and standard deviation $\sigma_\text{B}$,
then the likelihood of mean signal counts $n_\text{S}$ and background counts $n_\text{B}$ is that of the ``Gaussian-mean background'' problem \citep{cousins08} 
\begin{equation} \label{eq:gmb}
\begin{split}
   L(n_\text{S},n_\text{B}) =
&\frac{(n_\text{S}+n_\text{B})^{n_\text{on}}}{n_\text{on}!} e^{-(n_\text{S}+n_\text{B})} \\
&\times \frac{1}{\sqrt{2\pi \sigma_\text{B}^2}}e^{-\frac{1}{2}(n_\text{B}-\mu_\text{B})^2/\sigma_\text{B}^2}
\end{split}
\end{equation}
which describes the Poisson and Gaussian distributions underlying $n_\text{on}$ and $n_\text{B}$, respectively.

From the TBM, $\mu_\text{B} = \sum_i N_i T_i$ and $\sigma_\text{B}^2 = \sum_i (fN_i T_i)^2$, where the sum is over the number of observations, $N_i$ are the template scaling factors, $f$ is their fractional uncertainty (13\%), and $T_i$ are the counts predicted by the template, summed over the region of interest.
We note that the uncertainty relative to the mean therefore decreases as the square root of the number of observations, assuming that the uncertainty is dominated by independently-distributed, run-to-run uncertainties in the $N_i$ estimates.

From Wilks's theorem \citep{wilks38}, we can derive an expression for the significance
\begin{equation} \label{eq:gmbsig}
Z = \sqrt{2n_\text{on}(\ln(n_\text{on}/\hathat{n}_0)-1)  +  2\hathat{n}_0+(\hathat{n}_0-\mu_\text{B})^2/\sigma_\text{B}^2}
\end{equation}
where $\hathat{n}_0$ maximizes $L(0, \hathat{n}_0)$, found to be
\begin{equation}
    \hathat{n}_0 = \frac{1}{2}\left(\mu_\text{B}-\sigma_\text{B}^2\pm\sqrt{(\mu_\text{B}-\sigma_\text{B}^2)^2+4 n_\text{on}\sigma_\text{B}^2}\right)
\end{equation}
where only the positive solution is physical.

Upper limits on $n_\text{S}$ may be derived using the profile likelihood method (e.g. Refs. \cite{rolke05,cowan11,pdg20}), which corresponds asymptotically to the limits of \citet{feldman98}.
The limit is based on the likelihood ratio $\lambda$ given by
\begin{equation} \label{eq:lr}
\lambda = \frac{L(n_\text{S},\hathat{n}_\text{B})}{L(\hat{n}_\text{S},\hat{n}_\text{B})}
\end{equation}
where $\hathat{n}_\text{B}$ maximizes $L(n_\text{S},\hathat{n}_\text{B})$ for a given $n_\text{S}$, and $\hat{n}_\text{S}$ and $\hat{n}_\text{B}$ maximize $L$ without $n_\text{S}$ constrained.
Contours of constant $-2\ln \lambda$ correspond to different confidence levels on combinations of parameters.
For estimation of a single parameter $n_\text{S}$, $-2\ln \lambda=1.0$ corresponds to a 68.3\% confidence level, and $-2\ln \lambda=3.84$ to 95\% \cite{pdg20}.
We follow the method of \citet{rolke05}, wherein $n_\text{S}$ is increased until $-2\ln \lambda$ reaches a value corresponding to the desired confidence level, with $L(n_\text{S},\hathat{n}_\text{B})$ maximized at each $n_\text{S}$ when calculating $-2\ln \lambda$.
For the likelihood given by Equation \ref{eq:gmb}, $\hathat{n}_\text{B}$ for a given $n_\text{S}$ is
\begin{equation}
\hathat{n}_\text{B} = \frac{1}{2}\left(-n_\text{S}+\mu_\text{B}-\sigma_\text{B}^2\pm\sqrt{(n_\text{S}+\mu_\text{B}-\sigma_\text{B}^2)^2+4 n_\text{on}\sigma_\text{B}^2}\right)
\end{equation}
When considering the product of many likelihoods like Equation \ref{eq:gmb}, optimizations are instead performed numerically.

\subsubsection{Template Background Method Validation}
The validity of a background estimation method (and of the significance calculation) may be assessed by checking that the significances calculated at many trial positions over a field of view without gamma-ray sources are normally distributed (e.g. Ref. \cite{berge07}).
We perform this test on the Sgr A* Off data.
We note that the Sgr A* Off data are not identical to the templates due to the eight different observational pointing offsets.

\begin{figure}[t]
    \centering
    \includegraphics[width=0.5\textwidth]{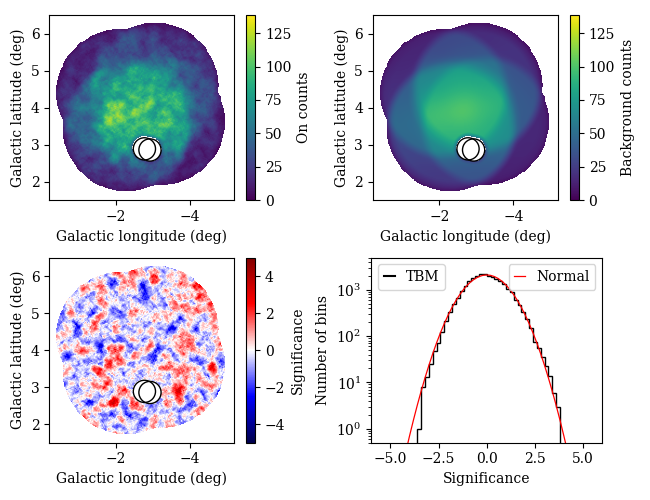}
\caption{
Sgr A* Off analysis using the template background method. 
The 2D maps show the integrated on counts (top left), estimated mean background counts (top right), and significance (bottom left).
Exclusion regions around the bright stars d Ophiuchi and HIP85442 (black) are indicated.
A histogram of significances is also shown (bottom right).
The red curve is the normal distribution, with unit variance and zero mean, multiplied by the total number of 2D significance map bins.
}
    \label{fig:tbmvalidation}
\end{figure}

We calculate significances in circular regions of $0.13\deg$ radius centered at each spatial bin with nonzero live time.
A 2D map and one-dimensional histogram of significances are shown in Figure \ref{fig:tbmvalidation}.
Maps of the observed counts and the predicted background counts are also shown.
The significances are approximately normally distributed, illustrating that the TBM provides an adequate description of the background over the entire field of view.

\section{Analysis} \label{sec:analysis}
We apply the TBM to the Sgr A* data, using the templates derived from the Sgr A* Off data.
The dark matter search region of interest is divided into three-dimensional bins in sky coordinates and energy, to improve the sensitivity of the search.
Significances are evaluated in each of these bins.
Fluxes (or flux limits) are measured and corresponding values of the dark matter thermally-averaged annihilation cross section are calculated, under assumptions for the dark matter halo profile and annihilation channel.

\subsection{Search Region}
Our dark matter search region of interest extends over a $2\deg$-radius circle centered at the GC.
This region is divided into multiple ring-shaped subregions to exploit the dark matter halo's spatial profile, as was done in similar analyses performed by the H.E.S.S. collaboration \citep{abdallah16,abdallah18,hess22}.
Sixteen spatial regions are defined, with inner radii $\theta$ from $0.4\deg$ to $1.9\deg$ and thickness $0.1\deg$ centered at $(\ell, b) = (0\deg, 0\deg)$.
Each subregion is also divided into 17 energy bins spaced logarithmically from 2 TeV to 100 TeV, 
corresponding to the upper 17 energy bins used in constructing the templates, which lie above our energy threshold of 2 TeV.
There are thus 272 bins in total.
The exclusion regions described in Section \ref{sec:data} are also used, and are shown in Figure \ref{fig:on}.
For each bin, we calculate $n_\text{on}$, $\mu_\text{B}$, $\sigma_\text{B}$, live time, and effective area.
The likelihood used for the calculations of upper limits is the product of the likelihoods for each bin.

\subsection{Gamma Rays from Dark Matter}
To connect the observed event counts to dark matter particle properties, assumptions must be made regarding the Galactic dark matter halo density profile and dark-matter-induced gamma-ray spectrum.
The expected number of signal counts $n_\text{S}$ within a solid angle $\Delta \Omega$ and energy range $\Delta E$, may be calculated for a given dark matter mass, annihilation channel, and thermally averaged cross section 
\begin{equation}
n_\text{S} =
\frac{1}{4\pi}\frac{\tac}{2\mdm^2} \int_{\Delta E}dE' \frac{dn}{dE'}  \overline A \int_{\Delta\Omega} d\Omega \; J  t_\text{live}
\end{equation}
where $\tac$ is the thermally-averaged dark matter annihilation cross section, $\mdm$ is the dark matter particle's mass, $dn/dE$ is the observed gamma-ray spectrum from a single dark matter annihilation, $\overline A$ is the energy-dependent effective area, $J$ is the J-factor, equal to the squared dark matter density integrated over a line of sight \citep{bergstrom98}, and $t_\text{live}$ is the live time for a position on the sky.
For each energy bin, an $\overline A$ is calculated by averaging the IRF-derived effective areas for each event $A_i$,
\begin{equation}
1/\overline A = \frac{1}{n_\text{on}}\sum_i^{n_\text{on}}1/A_i
\end{equation}
which accounts for the probability of observing each event being proportional to its $A_i$.

We define the J-factor
\begin{equation} \label{eq:jfactor}
J = \int_\text{los}d s \; \rho(r)^2 
\end{equation}
where $\rho(r)$ is the dark matter density a distance $r$ from the Galactic Center.
We note that this definition does not integrate over solid angle $\Omega$ and does not include multiplicative factors that make $J$ dimensionless.

We use an Einasto profile to model the Galactic dark matter halo density profile.
The Einasto profile is given by \citep{einasto65} 
\begin{equation}
\rho(r)=\rho_0 e^{-(2/\alpha)([r/r_s]^\alpha-1)}
\end{equation}
where $\rho_0$ is a density normalization factor, $r_s$ is a characteristic radius, and $\alpha$ influences the profile shape, yielding a cuspy profile for $\alpha<1$, and a cored profile otherwise.
We adopt parameters close to those used by previous Galactic Center gamma-ray dark matter studies \citep{abdallah16,acharyya21,hess22} to facilitate comparison, 
rather than using more recent values available in the literature.
We use $r_s=20$ kpc (also near to the values of e.g. Refs. \cite{pieri11,cirelli11}) and $\alpha=0.17$.
The normalization $\rho_0$ is chosen such that the local dark matter density is $0.4\text{ GeV cm}^{-3}$ at $r=8.5$ kpc.

For $dn/dE$ we consider individual annihilation channels, for the purpose of limit setting, and use the spectra of Ref. \cite{cirelli11}.
The energy resolution of our event reconstruction, $19\% \times E$, is incorporated by convolving $dn/dE \overline A$ with a Gaussian function with standard deviation $0.19E$.
The fluxes are multiplied by energy-dependent instrumental throughput factors before integrating over energy.

\section{Results} \label{sec:res}
\begin{figure}[t]
    \centering
    \includegraphics[width=0.5\textwidth]{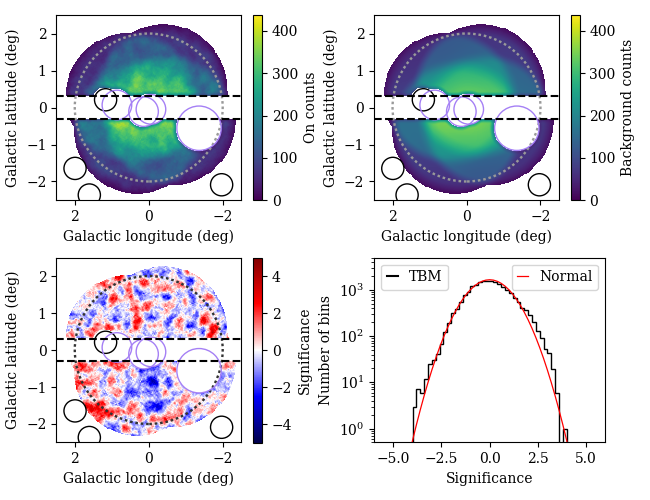}
    \caption{
Sgr A* analysis using the template background method. 
The 2D maps show the integrated on counts (top left), estimated mean background counts (top right), and significance (bottom left).
Exclusion regions around bright stars (black) and gamma ray sources (purple) are indicated.
The dashed black lines delineate $|b|<0.3\deg$.
Dotted lines encircle the $2\deg$-radius region of interest.
A histogram of significances is also shown (bottom right).
The red curve is the normal distribution, with unit variance and zero mean, multiplied by the total number of 2D significance map bins.
}
    \label{fig:on}
\end{figure}

The 2D significance map for our field of view around Sgr A* is shown in Figure \ref{fig:on}.
The significances are well-described by a normal distribution, suggesting that the background is adequately modeled.
The slight excess of positive significances corresponds mostly to spatial bins outside of our $2\deg$ region of interest.
No extended gamma-ray emission sources are indicated by the residuals.

\begin{figure}[t]
    \centering
    \includegraphics[width=0.5\textwidth]{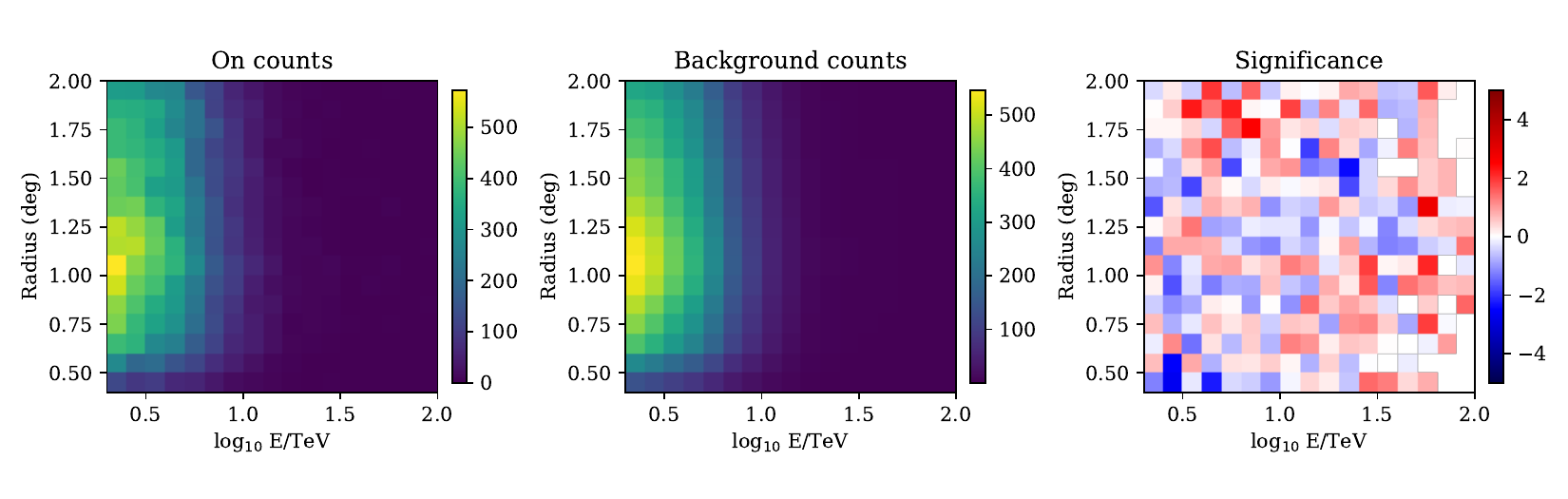}
    \caption{
On counts (left), expected background counts (middle), and significances (right), for each search region bin. Energy bins run along the $x$ axis and spatial bins run along the $y$ axis. No deviations from a background-only model are observed that exceed 3$\sigma$, even before trials factors are accounted for.
}
    \label{fig:roi}
\end{figure}

The results of the TBM for each search region bin are shown in Figure \ref{fig:roi}.
No significant excess is found in any dark matter search region, even without including trials factors for the number of search regions.
We therefore calculate upper limits on the dark matter annihilation cross section.

\begin{figure}[t]
    \centering
    \includegraphics[width=0.5\textwidth]{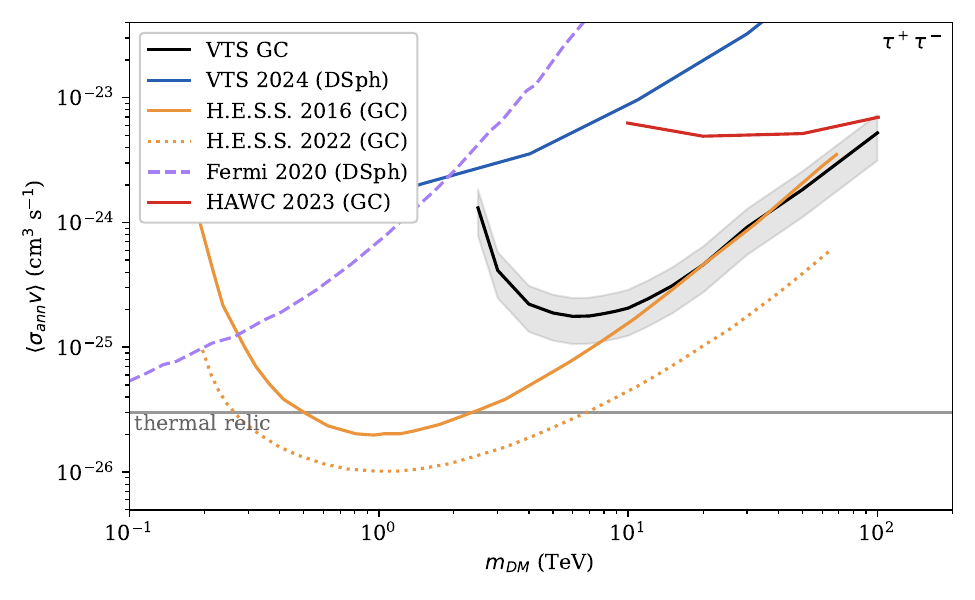} 
    \caption{
95\% confidence upper limit on $\tac$ assuming complete annihilation in the $\chi\chi \rightarrow \tau^+ \tau^-$ channel 
(black line), with 1$\sigma$ systematic uncertainty indicated (gray region).
Also shown are limits from the most recent VERITAS analysis of dwarf spheroidal galaxies  \citep{acharyya24}, as well as results from H.E.S.S. \citep{abdallah16,hess22}, {\fermi} \citep{hoof20}, and HAWC \citep{albert23}.
Limits assume Einasto profiles.
}
    \label{fig:limtau}
\end{figure}

\begin{figure}[t]
    \centering
    \includegraphics[width=0.5\textwidth]{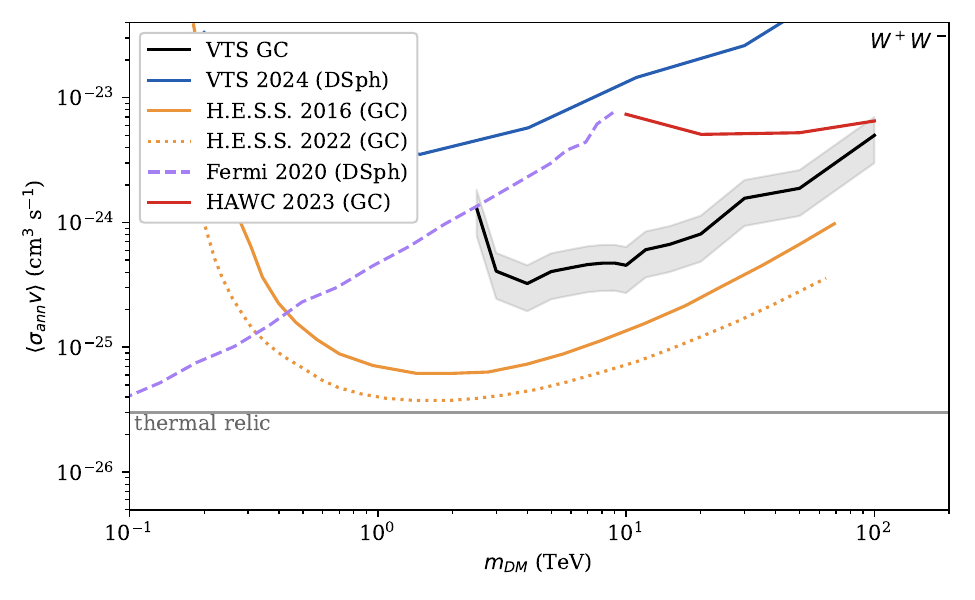}
    \caption{
Same as Figure \ref{fig:limtau}, but assuming complete annihilation in the $\chi\chi \rightarrow W^+ W^-$ channel.
}
    \label{fig:limw}
\end{figure}

\begin{figure}[t]
    \centering
    \includegraphics[width=0.5\textwidth]{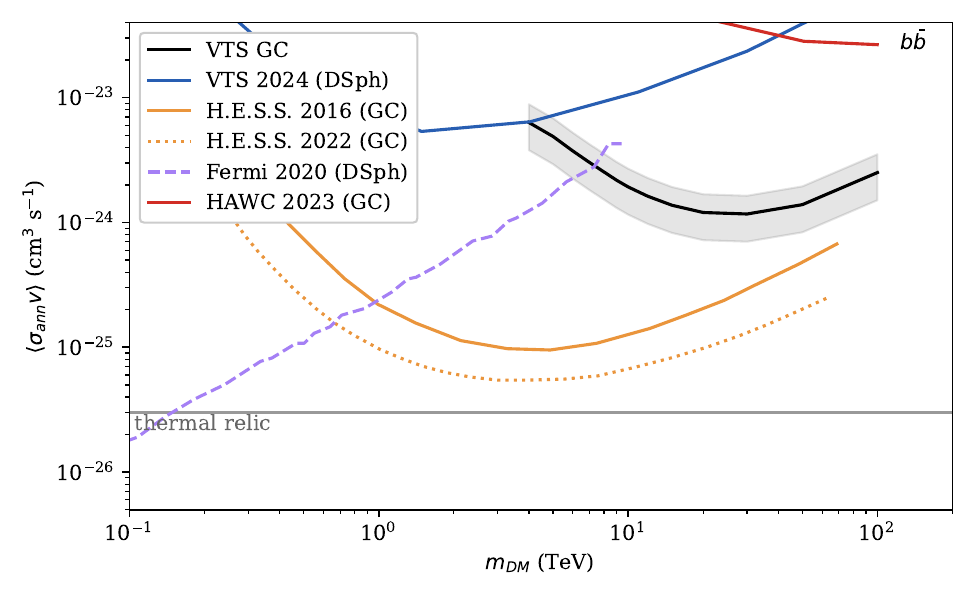}
    \caption{
Same as Figure \ref{fig:limtau}, but assuming complete annihilation in the $\chi\chi \rightarrow b \bar b$ channel.
}
    \label{fig:limb}
\end{figure}

We derive 95\% confidence level upper limits on $\tac$ for annihilation into the $\tau^+\tau^-$ final state in Figure \ref{fig:limtau}.
Similar limits are shown for $W^+W^-$ in Figure \ref{fig:limw} and $b \bar b$ in Figure \ref{fig:limb}.
The uncertainty due to the nominal VERITAS 40\% systematic uncertainty \cite{archer14} on LZA flux measurements is illustrated.
Limits from other experiments are also shown.

Our strongest constraints are in the $\tau^+\tau^-$ channel, where we obtain a 95\% confidence level upper limit $\tac < 1.77\times 10^{-25}\text{ cm}^3\text{ s}^{-1}$ at $\mdm = 6$ TeV for the Einasto profile.
This represents the most constraining dark matter limit from VERITAS.
Our constraints on $\tac$, reaching $\sim$$10^{-25}$--$10^{-24}\text{ cm}^3\text{ s}^{-1}$ in the 
$\tau^+\tau^-$, $W^+ W^-$, and $b \bar b$ channels, represent, together with the H.E.S.S. GC measurement \citep{hess22}, the strongest limits from any measurement to date for dark matter masses between 10 TeV and 100 TeV.
Our limit approaches that of Ref. \cite{hess22} despite the difference in live time (154.3 hr versus 546 hr) due to the effective area increase at LZA, resulting in a similar product of effective area and live time, $\sim$$2\times 10^{11} \text{ m}^2\text{ s}$ above 10 TeV near the GC.

\section{Discussion} \label{sec:disc}
The present work represents the strongest dark matter limits offered by VERITAS to date, due to the unique observational advantages of the Galactic Center region.
Our limits also illustrate the boosted effective areas at higher energies that are a consequence of LZA observations---limits derived from 178 hours of LZA data are comparable to those of Ref. \cite{abdallah16} with 254 hours.
While no dark matter-like signal is detected, we explore parameter space previously only reached by H.E.S.S. \citep{abdallah16,hess22}.
We remark that the TeV gamma-ray searches for dark matter remain most sensitive to photon, lepton, and charged-boson annihilation channels, and 
we reiterate that ample WIMP parameter space remains \citep{leane18}.
Some of this will be explored definitively by the Cherenkov Telescope Array Observatory (CTAO) \citep{acharyya21}.

Our background estimation technique, which is independent of gamma rays in the field of view, is subject to different systematics than previous searches that use traditional methods like reflected background regions.
The present approach is also sensitive to certain cored dark matter profiles, as well as extended, non-dark matter gamma-ray emission.
In the present analysis no extended sources are found, however.

\section{Conclusions} \label{sec:conc}
We have carried out a sensitive dark matter search using 177.8 hours of gamma-ray observations of the Galactic Center taken by VERITAS over a 12 year period.
A new background estimation method was developed to facilitate this search, and its validity was demonstrated.
No significant gamma-ray excess was found,
enabling strong limits to be placed on the dark matter annihilation cross section, reaching as low as $\tac < 1.77\times 10^{-25}\text{ cm}^3\text{ s}^{-1}$ at 6 TeV in the $\tau^+\tau^-$ channel.
There remain methodological improvements that could improve these limits, such as more efficient gamma-hadron separation, or simultaneous modeling of dark matter and astrophysical source emission including Sgr A*, where the dark matter J-factor also peaks.
Beyond this, progress relies on instrumental upgrades like those offered by CTAO.

\section{Acknowledgments} \label{sec:ack}
This research is supported by grants from the U.S. Department of Energy Office of Science, the U.S. National Science Foundation and the Smithsonian Institution, by NSERC in Canada, and by the Helmholtz Association in Germany. This research used resources provided by the Open Science Grid, which is supported by the National Science Foundation and the U.S. Department of Energy's Office of Science, and resources of the National Energy Research Scientific Computing Center (NERSC), a U.S. Department of Energy Office of Science User Facility operated under Contract No. DE-AC02-05CH11231. We acknowledge the excellent work of the technical support staff at the Fred Lawrence Whipple Observatory and at the collaborating institutions in the construction and operation of VERITAS.


\bibliographystyle{apsrev4-2prd}
\bibliography{vtsdm}

@ARTICLE{abe23,
  title = {Search for Gamma-Ray Spectral Lines from Dark Matter Annihilation up to 100 TeV toward the Galactic Center with MAGIC},
  author = {Abe, H. and Abe, S. and Acciari, V. A. and Aniello, T. and Ansoldi, S. and Antonelli, L. A. and Arbet Engels, A. and Arcaro, C. and Artero, M. and Asano, K. and Baack, D. and Babi\ifmmode \acute{c}\else \'{c}\fi{}, A. and Baquero, A. and Barres de Almeida, U. and Barrio, J. A. and Batkovi\ifmmode \acute{c}\else \'{c}\fi{}, I. and Baxter, J. and Becerra Gonz\'alez, J. and Bednarek, W. and Bernardini, E. and Bernardos, M. and Berti, A. and Besenrieder, J. and Bhattacharyya, W. and Bigongiari, C. and Biland, A. and Blanch, O. and Bonnoli, G. and Bo\ifmmode \check{s}\else \v{s}\fi{}njak, \ifmmode \check{Z}\else \v{Z}\fi{}. and Burelli, I. and Busetto, G. and Carosi, R. and Carretero-Castrillo, M. and Ceribella, G. and Chai, Y. and Chilingarian, A. and Cikota, S. and Colombo, E. and Contreras, J. L. and Cortina, J. and Covino, S. and D'Amico, G. and D'Elia, V. and Da Vela, P. and Dazzi, F. and De Angelis, A. and De Lotto, B. and Del Popolo, A. and Delfino, M. and Delgado, J. and Delgado Mendez, C. and Depaoli, D. and Di Pierro, F. and Di Venere, L. and Do Souto Espi\~neira, E. and Dominis Prester, D. and Donini, A. and Dorner, D. and Doro, M. and Elsaesser, D. and Emery, G. and Fallah Ramazani, V. and Fari\~na, L. and Fattorini, A. and Font, L. and Fruck, C. and Fukami, S. and Fukazawa, Y. and Garc\'{\i}a L\'opez, R. J. and Garczarczyk, M. and Gasparyan, S. and Gaug, M. and Giesbrecht Paiva, J. G. and Giglietto, N. and Giordano, F. and Gliwny, P. and Godinovi\ifmmode \acute{c}\else \'{c}\fi{}, N. and Green, J. G. and Green, D. and Hadasch, D. and Hahn, A. and Hassan, T. and Heckmann, L. and Herrera, J. and Hrupec, D. and H\"utten, M. and Imazawa, R. and Inada, T. and Iotov, R. and Ishio, K. and Jim\'enez Mart\'{\i}nez, I. and Jormanainen, J. and Kerszberg, D. and Kobayashi, Y. and Kubo, H. and Kushida, J. and Lamastra, A. and Lelas, D. and Leone, F. and Lindfors, E. and Linhoff, L. and Lombardi, S. and Longo, F. and L\'opez-Coto, R. and L\'opez-Moya, M. and L\'opez-Oramas, A. and Loporchio, S. and Lorini, A. and Lyard, E. and Machado de Oliveira Fraga, B. and Majumdar, P. and Makariev, M. and Maneva, G. and Mang, N. and Manganaro, M. and Mangano, S. and Mannheim, K. and Mariotti, M. and Mart\'{\i}nez, M. and Mas Aguilar, A. and Mazin, D. and Menchiari, S. and Mender, S. and Mi\ifmmode \acute{c}\else \'{c}\fi{}anovi\ifmmode \acute{c}\else \'{c}\fi{}, S. and Miceli, D. and Miener, T. and Miranda, J. M. and Mirzoyan, R. and Molina, E. and Mondal, H. A. and Moralejo, A. and Morcuende, D. and Moreno, V. and Nakamori, T. and Nanci, C. and Nava, L. and Neustroev, V. and Nievas Rosillo, M. and Nigro, C. and Nilsson, K. and Nishijima, K. and Njoh Ekoume, T. and Noda, K. and Nozaki, S. and Ohtani, Y. and Oka, T. and Otero-Santos, J. and Paiano, S. and Palatiello, M. and Paneque, D. and Paoletti, R. and Paredes, J. M. and Pavleti\ifmmode \acute{c}\else \'{c}\fi{}, L. and Persic, M. and Pihet, M. and Podobnik, F. and Prada Moroni, P. G. and Prandini, E. and Principe, G. and Priyadarshi, C. and Puljak, I. and Rhode, W. and Rib\'o, M. and Rico, J. and Righi, C. and Rugliancich, A. and Sahakyan, N. and Saito, T. and Sakurai, S. and Satalecka, K. and Saturni, F. G. and Schleicher, B. and Schmidt, K. and Schmuckermaier, F. and Schubert, J. L. and Schweizer, T. and Sitarek, J. and Sliusar, V. and Sobczynska, D. and Spolon, A. and Stamerra, A. and Stri\ifmmode \check{s}\else \v{s}\fi{}kovi\ifmmode \acute{c}\else \'{c}\fi{}, J. and Strom, D. and Strzys, M. and Suda, Y. and Suri\ifmmode \acute{c}\else \'{c}\fi{}, T. and Takahashi, M. and Takeishi, R. and Tavecchio, F. and Temnikov, P. and Terauchi, K. and Terzi\ifmmode \acute{c}\else \'{c}\fi{}, T. and Teshima, M. and Tosti, L. and Truzzi, S. and Tutone, A. and Ubach, S. and van Scherpenberg, J. and Vazquez Acosta, M. and Ventura, S. and Verguilov, V. and Viale, I. and Vigorito, C. F. and Vitale, V. and Vovk, I. and Walter, R. and Will, M. and Wunderlich, C. and Yamamoto, T. and Zari\ifmmode \acute{c}\else \'{c}\fi{}, D. and Hiroshima, N. and Kohri, K.},
  collaboration = {MAGIC Collaboration},
  journal = {Phys. Rev. Lett.},
  volume = {130},
  issue = {6},
  pages = {061002},
  numpages = {8},
  year = {2023},
  month = {Feb},
  publisher = {American Physical Society},
  doi = {10.1103/PhysRevLett.130.061002},
  url = {https://link.aps.org/doi/10.1103/PhysRevLett.130.061002}
}

@ARTICLE{abdallah18,
  title = {Search for $\ensuremath{\gamma}$-Ray Line Signals from Dark Matter Annihilations in the Inner Galactic Halo from 10 Years of Observations with H.E.S.S.},
  author = {Abdallah, H. and Abramowski, A. and Aharonian, F. and Ait Benkhali, F. and Ang\"uner, E. O. and Arakawa, M. and Arrieta, M. and Aubert, P. and Backes, M. and Balzer, A. and Barnard, M. and Becherini, Y. and Becker Tjus, J. and Berge, D. and Bernhard, S. and Bernl\"ohr, K. and Blackwell, R. and B\"ottcher, M. and Boisson, C. and Bolmont, J. and Bonnefoy, S. and Bordas, P. and Bregeon, J. and Brun, F. and Brun, P. and Bryan, M. and B\"uchele, M. and Bulik, T. and Capasso, M. and Caroff, S. and Carosi, A. and Carr, J. and Casanova, S. and Cerruti, M. and Chakraborty, N. and Chaves, R. C. G. and Chen, A. and Chevalier, J. and Colafrancesco, S. and Condon, B. and Conrad, J. and Davids, I. D. and Decock, J. and Deil, C. and Devin, J. and deWilt, P. and Dirson, L. and Djannati-Ata\"{\i}, A. and Domainko, W. and Donath, A. and Drury, L. O'C. and Dutson, K. and Dyks, J. and Edwards, T. and Egberts, K. and Eger, P. and Emery, G. and Ernenwein, J.-P. and Eschbach, S. and Farnier, C. and Fegan, S. and Fernandes, M. V. and Fiasson, A. and Fontaine, G. and F\"orster, A. and Funk, S. and F\"u\ss{}ling, M. and Gabici, S. and Gallant, Y. A. and Garrigoux, T. and Gat\'e, F. and Giavitto, G. and Giebels, B. and Glawion, D. and Glicenstein, J. F. and Gottschall, D. and Grondin, M.-H. and Hahn, J. and Haupt, M. and Hawkes, J. and Heinzelmann, G. and Henri, G. and Hermann, G. and Hinton, J. A. and Hofmann, W. and Hoischen, C. and Holch, T. L. and Holler, M. and Horns, D. and Ivascenko, A. and Iwasaki, H. and Jacholkowska, A. and Jamrozy, M. and Janiak, M. and Jankowsky, D. and Jankowsky, F. and Jingo, M. and Jouvin, L. and Jung-Richardt, I. and Kastendieck, M. A. and Katarzy\ifmmode \acute{n}\else \'{n}\fi{}ski, K. and Katsuragawa, M. and Katz, U. and Kerszberg, D. and Khangulyan, D. and Kh\'elifi, B. and King, J. and Klepser, S. and Klochkov, D. and Klu\ifmmode \acute{z}\else \'{z}\fi{}niak, W. and Komin, Nu. and Kosack, K. and Krakau, S. and Kraus, M. and Kr\"uger, P. P. and Laffon, H. and Lamanna, G. and Lau, J. and Lees, J.-P. and Lefaucheur, J. and Lemi\`ere, A. and Lemoine-Goumard, M. and Lenain, J.-P. and Leser, E. and Liu, R. and Lohse, T. and Lorentz, M. and L\'opez-Coto, R. and Lypova, I. and Malyshev, D. and Marandon, V. and Marcowith, A. and Mariaud, C. and Marx, R. and Maurin, G. and Maxted, N. and Mayer, M. and Meintjes, P. J. and Meyer, M. and Mitchell, A. M. W. and Moderski, R. and Mohamed, M. and Mohrmann, L. and Mor\aa{}, K. and Moulin, E. and Murach, T. and Nakashima, S. and de Naurois, M. and Ndiyavala, H. and Niederwanger, F. and Niemiec, J. and Oakes, L. and O'Brien, P. and Odaka, H. and Ohm, S. and Ostrowski, M. and Oya, I. and Padovani, M. and Panter, M. and Parsons, R. D. and Pekeur, N. W. and Pelletier, G. and Perennes, C. and Petrucci, P.-O. and Peyaud, B. and Piel, Q. and Pita, S. and Poireau, V. and Poon, H. and Prokhorov, D. and Prokoph, H. and P\"uhlhofer, G. and Punch, M. and Quirrenbach, A. and Raab, S. and Rauth, R. and Reimer, A. and Reimer, O. and Renaud, M. and de los Reyes, R. and Rieger, F. and Rinchiuso, L. and Romoli, C. and Rowell, G. and Rudak, B. and Rulten, C. B. and Sahakian, V. and Saito, S. and Sanchez, D. A. and Santangelo, A. and Sasaki, M. and Schandri, M. and Schlickeiser, R. and Sch\"ussler, F. and Schulz, A. and Schwanke, U. and Schwemmer, S. and Seglar-Arroyo, M. and Settimo, M. and Seyffert, A. S. and Shafi, N. and Shilon, I. and Shiningayamwe, K. and Simoni, R. and Sol, H. and Spanier, F. and Spir-Jacob, M. and Stawarz, \L{}. and Steenkamp, R. and Stegmann, C. and Steppa, C. and Sushch, I. and Takahashi, T. and Tavernet, J.-P. and Tavernier, T. and Taylor, A. M. and Terrier, R. and Tibaldo, L. and Tiziani, D. and Tluczykont, M. and Trichard, C. and Tsirou, M. and Tsuji, N. and Tuffs, R. and Uchiyama, Y. and van der Walt, J. and van Eldik, C. and van Rensburg, C. and van Soelen, B. and Vasileiadis, G. and Veh, J. and Venter, C. and Viana, A. and Vincent, P. and Vink, J. and Voisin, F. and V\"olk, H. J. and Vuillaume, T. and Wadiasingh, Z. and Wagner, S. J. and Wagner, P. and Wagner, R. M. and White, R. and Wierzcholska, A. and Willmann, P. and W\"ornlein, A. and Wouters, D. and Yang, R. and Zaborov, D. and Zacharias, M. and Zanin, R. and Zdziarski, A. A. and Zech, A. and Zefi, F. and Ziegler, A. and Zorn, J. and \ifmmode \dot{Z}\else \.{Z}\fi{}ywucka, N.},
  collaboration = {H.E.S.S. Collaboration},
  journal = {Phys. Rev. Lett.},
  volume = {120},
  issue = {20},
  pages = {201101},
  numpages = {7},
  year = {2018},
  month = {May},
  publisher = {American Physical Society},
  doi = {10.1103/PhysRevLett.120.201101},
  url = {https://link.aps.org/doi/10.1103/PhysRevLett.120.201101}
}

@ARTICLE{acharyya21,
doi = {10.1088/1475-7516/2021/01/057},
url = {https://doi.org/10.1088/1475-7516/2021/01/057},
year = {2021},
month = {jan},
publisher = {},
volume = {2021},
number = {01},
pages = {057},
author = {Acharyya, A. and Adam, R. and Adams, C. and Agudo, I. and Aguirre-Santaella, A. and Alfaro, R. and Alfaro, J. and Alispach, C. and Aloisio, R. and Batista, R. Alves and Amati, L. and Ambrosi, G. and Angüner, E.O. and Antonelli, L.A. and Aramo, C. and Araudo, A. and Armstrong, T. and Arqueros, F. and Asano, K. and Ascasíbar, Y. and Ashley, M. and Balazs, C. and Ballester, O. and Larriva, A. Baquero and Martins, V. Barbosa and Barkov, M. and de Almeida, U. Barres and Barrio, J.A. and Bastieri, D. and Becerra, J. and Beck, G. and Tjus, J. Becker and Benbow, W. and Benito, M. and Berge, D. and Bernardini, E. and Bernlöhr, K. and Berti, A. and Bertucci, B. and Beshley, V. and Biasuzzi, B. and Biland, A. and Bissaldi, E. and Biteau, J. and Blanch, O. and Blazek, J. and Bocchino, F. and Boisson, C. and Arbeletche, L. Bonneau and Bordas, P. and Bosnjak, Z. and Bottacini, E. and Bozhilov, V. and Bregeon, J. and Brill, A. and Bringmann, T. and Brown, A.M. and Brun, P. and Brun, F. and Bruno, P. and Bulgarelli, A. and Burton, M. and Burtovoi, A. and Buscemi, M. and Cameron, R. and Capasso, M. and Caproni, A. and Capuzzo-Dolcetta, R. and Caraveo, P. and Carosi, R. and Carosi, A. and Casanova, S. and Cascone, E. and Cassol, F. and Catalani, F. and Cauz, D. and Cerruti, M. and Chadwick, P. and Chaty, S. and Chen, A. and Chernyakova, M. and Chiaro, G. and Chiavassa, A. and Chikawa, M. and Chudoba, J. and Çolak, M. and Conforti, V. and Coniglione, R. and Conte, F. and Contreras, J.L. and Coronado-Blazquez, J. and Costa, A. and Costantini, H. and Cotter, G. and Cristofari, P. and D'A𝚤, A. and D'Ammando, F. and Damone, L.A. and Daniel, M.K. and Dazzi, F. and Angelis, A. De and Caprio, V. De and Anjos, R. de Cássia dos and Pino, E.M. de Gouveia Dal and Lotto, B. De and Martino, D. De and Wilhelmi, E. de Oña and Palma, F. De and de Souza, V. and Delgado, C. and Giler, A.G. Delgado and Volpe, D. della and Depaoli, D. and Girolamo, T. Di and Pierro, F. Di and Venere, L. Di and Diebold, S. and Dmytriiev, A. and Domínguez, A. and Donini, A. and Doro, M. and Ebr, J. and Eckner, C. and Edwards, T. D. P. and Ekoume, T.R.N. and Elsässer, D. and Evoli, C. and Falceta-Goncalves, D. and Fedorova, E. and Fegan, S. and Feng, Q. and Ferrand, G. and Ferrara, G. and Fiandrini, E. and Fiasson, A. and Filipovic, M. and Fioretti, V. and Fiori, M. and Foffano, L. and Fontaine, G. and Fornieri, O. and Franco, F.J. and Fukami, S. and Fukui, Y. and Gaggero, D. and Galaz, G. and Gammaldi, V. and Garcia, E. and Garczarczyk, M. and Gascon, D. and Gent, A. and Ghalumyan, A. and Gianotti, F. and Giarrusso, M. and Giavitto, G. and Giglietto, N. and Giordano, F. and Giuliani, A. and Glicenstein, J. and Gnatyk, R. and Goldoni, P. and González, M.M. and Gourgouliatos, K. and Granot, J. and Grasso, D. and Green, J. and Grillo, A. and Gueta, O. and Gunji, S. and Halim, A. and Hassan, T. and Heller, M. and Cadena, S. Hernández and Hiroshima, N. and Hnatyk, B. and Hofmann, W. and Holder, J. and Horan, D. and Hörandel, J. and Horvath, P. and Hovatta, T. and Hrabovsky, M. and Hrupec, D. and Hughes, G. and Humensky, T.B. and Hütten, M. and Iarlori, M. and Inada, T. and Inoue, S. and Iocco, F. and Iori, M. and Jamrozy, M. and Janecek, P. and Jin, W. and Jouvin, L. and Jurysek, J. and Karukes, E. and Katarzyński, K. and Kazanas, D. and Kerszberg, D. and Kherlakian, M.C. and Kissmann, R. and Knödlseder, J. and Kobayashi, Y. and Kohri, K. and Komin, N. and Kubo, H. and Kushida, J. and Lamanna, G. and Lapington, J. and Laporte, P. and de Oliveira, M.A. Leigui and Lenain, J. and Leone, F. and Leto, G. and Lindfors, E. and Lohse, T. and Lombardi, S. and Longo, F. and Lopez, A. and López, M. and López-Coto, R. and Loporchio, S. and Luque-Escamilla, P.L. and Mach, E. and Maggio, C. and Maier, G. and Mallamaci, M. and de Almeida, R. Malta Nunes and Mandat, D. and Manganaro, M. and Mangano, S. and Manicò, G. and Marculewicz, M. and Mariotti, M. and Markoff, S. and Marquez, P. and Martí, J. and Martinez, O. and Martínez, M. and Martínez, G. and Martínez-Huerta, H. and Maurin, G. and Mazin, D. and Mbarubucyeye, J.D. and Miranda, D. Medina and Meyer, M. and Miceli, M. and Miener, T. and Minev, M. and Miranda, J.M. and Mirzoyan, R. and Mizuno, T. and Mode, B. and Moderski, R. and Mohrmann, L. and Molina, E. and Montaruli, T. and Moralejo, A. and Morcuende-Parrilla, D. and Morselli, A. and Mukherjee, R. and Mundell, C. and Nagai, A. and Nakamori, T. and Nemmen, R. and Niemiec, J. and Nieto, D. and Nikołajuk, M. and Ninci, D. and Noda, K. and Nosek, D. and Nozaki, S. and Ohira, Y. and Ohishi, M. and Ohtani, Y. and Oka, T. and Okumura, A. and Ong, R.A. and Orienti, M. and Orito, R. and Orlandini, M. and Orlando, S. and Orlando, E. and Ostrowski, M. and Oya, I. and Pagano, I. and Pagliaro, A. and Palatiello, M. and Pantaleo, F.R. and Paredes, J.M. and Pareschi, G. and Parmiggiani, N. and Patricelli, B. and Pavletić, L. and Pe'er, A. and Pecimotika, M. and Pérez-Romero, J. and Persic, M. and Petruk, O. and Pfrang, K. and Piano, G. and Piatteli, P. and Pietropaolo, E. and Pillera, R. and Pilszyk, B. and Pintore, F. and Pohl, M. and Poireau, V. and Prado, R.R. and Prandini, E. and Prast, J. and Principe, G. and Prokoph, H. and Prouza, M. and Przybilski, H. and Pühlhofer, G. and Pumo, M.L. and Queiroz, F. and Quirrenbach, A. and Rainò, S. and Rando, R. and Razzaque, S. and Recchia, S. and Reimer, O. and Reisenegger, A. and Renier, Y. and Rhode, W. and Ribeiro, D. and Ribó, M. and Richtler, T. and Rico, J. and Rieger, F. and Rinchiuso, L. and Rizi, V. and Rodriguez, J. and Fernandez, G. Rodriguez and Ramirez, J.C. Rodriguez and Rojas, G. and Romano, P. and Romeo, G. and Rosado, J. and Rowell, G. and Rudak, B. and Russo, F. and Sadeh, I. and Sæther Hatlen, E. and Safi-Harb, S. and Greus, F. Salesa and Salina, G. and Sanchez, D. and Sánchez-Conde, M. and Sangiorgi, P. and Sano, H. and Santander, M. and Santos, E.M. and Santos-Lima, R. and Sanuy, A. and Sarkar, S. and Saturni, F.G. and Sawangwit, U. and Schussler, F. and Schwanke, U. and Sciacca, E. and Scuderi, S. and Seglar-Arroyo, M. and Sergijenko, O. and Servillat, M. and Seweryn, K. and Shalchi, A. and Sharma, P. and Shellard, R.C. and Siejkowski, H. and Silk, J. and Siqueira, C. and Sliusar, V. and Słowikowska, A. and Sokolenko, A. and Sol, H. and Spencer, S. and Stamerra, A. and Stanič, S. and Starling, R. and Stolarczyk, T. and Straumann, U. and Strišković, J. and Suda, Y. and Suomijarvi, T. and Świerk, P. and Tavecchio, F. and Taylor, L. and Tejedor, L.A. and Teshima, M. and Testa, V. and Tibaldo, L. and Peixoto, C.J. Todero and Tokanai, F. and Tonev, D. and Tosti, G. and Tosti, L. and Tothill, N. and Truzzi, S. and Travnicek, P. and Vagelli, V. and Vallage, B. and Vallania, P. and Eldik, C. van and Vandenbroucke, J. and Varner, G.S. and Vassiliev, V. and Acosta, M. Vázquez and Vecchi, M. and Ventura, S. and Vercellone, S. and Vergani, S. and Verna, G. and Viana, A. and Vigorito, C.F. and Vink, J. and Vitale, V. and Vorobiov, S. and Vovk, I. and Vuillaume, T. and Wagner, S.J. and Walter, R. and Watson, J. and Weniger, C. and White, R. and White, M. and Wiemann, R. and Wierzcholska, A. and Will, M. and Williams, D.A. and Wischnewski, R. and Yanagita, S. and Yang, L. and Yoshikoshi, T. and Zacharias, M. and Zaharijas, G. and Zakaria, A.A. and Zampieri, L. and Zanin, R. and Zaric, D. and Zavrtanik, M. and Zavrtanik, D. and Zdziarski, A.A. and Zech, A. and Zechlin, H. and Zhdanov, V.I. and Živec, M.},
title = {Sensitivity of the Cherenkov Telescope Array to a dark matter signal from the Galactic centre},
journal = {\jcap}
}

@ARTICLE{acharyya24,
  title = {Indirect search for dark matter with a combined analysis of dwarf spheroidal galaxies from VERITAS},
  author = {Acharyya, A. and Adams, C. B. and Bangale, P. and Bartkoske, J. T. and Batista, P. and Benbow, W. and Christiansen, J. L. and Chromey, A. J. and Duerr, A. and Errando, M. and Falcone, A. and Feng, Q. and Foote, G. M. and Fortson, L. and Furniss, A. and Hanlon, W. and Hanna, D. and Hervet, O. and Hinrichs, C. E. and Holder, J. and Humensky, T. B. and Jin, W. and Johnson, M. N. and Kaaret, P. and Kertzman, M. and Kieda, D. and Kleiner, T. K. and Korzoun, N. and Kumar, S. and Lang, M. J. and Lundy, M. and Maier, G. and McGrath, C. E. and Millard, M. J. and Mooney, C. L. and Moriarty, P. and Mukherjee, R. and Ning, W. and O'Brien, S. and Ong, R. A. and Park, N. and Pohl, M. and Pueschel, E. and Quinn, J. and Rabinowitz, P. L. and Ragan, K. and Reynolds, P. T. and Ribeiro, D. and Roache, E. and Ryan, J. L. and Sadeh, I. and Saha, L. and Sembroski, G. H. and Shang, R. and Splettstoesser, M. and Tak, D. and Talluri, A. K. and Tucci, J. V. and Vassiliev, V. V. and Weinstein, A. and Williams, D. A. and Wong, S. L.},
  collaboration = {VERITAS Collaboration},
  journal = {Phys. Rev. D},
  volume = {110},
  issue = {6},
  pages = {063034},
  numpages = {14},
  year = {2024},
  month = {Sep},
  publisher = {American Physical Society},
  doi = {10.1103/PhysRevD.110.063034},
  url = {https://link.aps.org/doi/10.1103/PhysRevD.110.063034}
}

@ARTICLE{ackermann17,
doi = {10.3847/1538-4357/aa6cab},
url = {https://doi.org/10.3847/1538-4357/aa6cab},
year = {2017},
month = {may},
publisher = {The American Astronomical Society},
volume = {840},
number = {1},
pages = {43},
author = {Ackermann, M. and Ajello, M. and Albert, A. and Atwood, W. B. and Baldini, L. and Ballet, J. and Barbiellini, G. and Bastieri, D. and Bellazzini, R. and Bissaldi, E. and Blandford, R. D. and Bloom, E. D. and Bonino, R. and Bottacini, E. and Brandt, T. J. and Bregeon, J. and Bruel, P. and Buehler, R. and Burnett, T. H. and Cameron, R. A. and Caputo, R. and Caragiulo, M. and Caraveo, P. A. and Cavazzuti, E. and Cecchi, C. and Charles, E. and Chekhtman, A. and Chiang, J. and Chiappo, A. and Chiaro, G. and Ciprini, S. and Conrad, J. and Costanza, F. and Cuoco, A. and Cutini, S. and D’Ammando, F. and Palma, F. de and Desiante, R. and Digel, S. W. and Lalla, N. Di and Mauro, M. Di and Venere, L. Di and Drell, P. S. and Favuzzi, C. and Fegan, S. J. and Ferrara, E. C. and Focke, W. B. and Franckowiak, A. and Fukazawa, Y. and Funk, S. and Fusco, P. and Gargano, F. and Gasparrini, D. and Giglietto, N. and Giordano, F. and Giroletti, M. and Glanzman, T. and Gomez-Vargas, G. A. and Green, D. and Grenier, I. A. and Grove, J. E. and Guillemot, L. and Guiriec, S. and Gustafsson, M. and Harding, A. K. and Hays, E. and Hewitt, J. W. and Horan, D. and Jogler, T. and Johnson, A. S. and Kamae, T. and Kocevski, D. and Kuss, M. and Mura, G. La and Larsson, S. and Latronico, L. and Li, J. and Longo, F. and Loparco, F. and Lovellette, M. N. and Lubrano, P. and Magill, J. D. and Maldera, S. and Malyshev, D. and Manfreda, A. and Martin, P. and Mazziotta, M. N. and Michelson, P. F. and Mirabal, N. and Mitthumsiri, W. and Mizuno, T. and Moiseev, A. A. and Monzani, M. E. and Morselli, A. and Negro, M. and Nuss, E. and Ohsugi, T. and Orienti, M. and Orlando, E. and Ormes, J. F. and Paneque, D. and Perkins, J. S. and Persic, M. and Pesce-Rollins, M. and Piron, F. and Principe, G. and Rainò, S. and Rando, R. and Razzano, M. and Razzaque, S. and Reimer, A. and Reimer, O. and Sánchez-Conde, M. and Sgrò, C. and Simone, D. and Siskind, E. J. and Spada, F. and Spandre, G. and Spinelli, P. and Suson, D. J. and Tajima, H. and Tanaka, K. and Thayer, J. B. and Tibaldo, L. and Torres, D. F. and Troja, E. and Uchiyama, Y. and Vianello, G. and Wood, K. S. and Wood, M. and Zaharijas, G. and Zimmer, S. and (The Fermi LAT Collaboration)},
title = {The Fermi Galactic Center GeV Excess and Implications for Dark Matter},
journal = {\apj},
}

@ARTICLE{adams21,
doi = {10.3847/1538-4357/abf926},
url = {https://doi.org/10.3847/1538-4357/abf926},
year = {2021},
month = {jun},
publisher = {The American Astronomical Society},
volume = {913},
number = {2},
pages = {115},
author = {Adams, C. B. and Benbow, W. and Brill, A. and Brose, R. and Buchovecky, M. and Capasso, M. and Christiansen, J. L. and Chromey, A. J. and Daniel, M. K. and Errando, M. and Falcone, A. and Feng, Q. and Finley, J. P. and Fortson, L. and Furniss, A. and Gent, A. and Gillanders, G. H. and Giuri, C. and Hanna, D. and Hervet, O. and Holder, J. and Hughes, G. and Humensky, T. B. and Jin, W. and Kaaret, P. and Kelley-Hoskins, N. and Kertzman, M. and Kieda, D. and Krennrich, F. and Kumar, S. and Lang, M. J. and Lundy, M. and Maier, G. and Moriarty, P. and Mukherjee, R. and Nieto, D. and Nievas-Rosillo, M. and O’Brien, S. and Ong, R. A. and Otte, A. N. and Pfrang, K. and Pohl, M. and Prado, R. R. and Pueschel, E. and Quinn, J. and Ragan, K. and Reynolds, P. T. and Ribeiro, D. and Richards, G. T. and Roache, E. and Ryan, J. L. and Santander, M. and Schlenstedt, S. and Sembroski, G. H. and Shang, R. and Stevenson, B. and Wakely, S. P. and Weinstein, A. and Williams, D. A.},
title = {VERITAS Observations of the Galactic Center Region at Multi-TeV Gamma-Ray Energies},
journal = {\apj}
}

@ARTICLE{adams22,
       author = {{Adams}, C.~B. and {Benbow}, W. and {Brill}, A. and {Buckley}, J.~H. and {Christiansen}, J.~L. and {Falcone}, A. and {Feng}, Q. and {Finley}, J.~P. and {Foote}, Juniper and {Fortson}, L. and {Furniss}, A. and {Giuri}, C. and {Hanna}, D. and {Hassan}, T. and {Hervet}, O. and {Holder}, J. and {Hona}, B. and {Humensky}, T.~B. and {Jin}, W. and {Kaaret}, P. and {Kleiner}, T.~K. and {Kumar}, S. and {Lang}, M.~J. and {Lundy}, M. and {Maier}, G. and {Moriarty}, P. and {Mukherjee}, R. and {Nievas Rosillo}, M. and {O'Brien}, S. and {Park}, N. and {Patel}, S. and {Pfrang}, K. and {Pohl}, M. and {Prado}, R.~R. and {Pueschel}, E. and {Quinn}, J. and {Ragan}, K. and {Reynolds}, P.~T. and {Ribeiro}, D. and {Roache}, E. and {Ryan}, J.~L. and {Santander}, M. and {Weinstein}, A. and {Williams}, D.~A. and {Williamson}, T.~J.},
        title = "{The throughput calibration of the VERITAS telescopes}",
      journal = {\aap},
         year = 2022,
        month = feb,
       volume = {658},
          eid = {A83},
        pages = {A83},
          doi = {10.1051/0004-6361/202142275},
       url = {https://doi.org/10.1051/0004-6361/202142275}
}

@ARTICLE{albert23,
doi = {10.1088/1475-7516/2023/12/038},
url = {https://doi.org/10.1088/1475-7516/2023/12/038},
year = {2023},
month = {dec},
publisher = {IOP Publishing},
volume = {2023},
number = {12},
pages = {038},
author = {Albert, A. and Alfaro, R. and Alvarez, C. and Arteaga-Velázquez, J.C. and Avila Rojas, D. and Ayala Solares, H.A. and Belmont-Moreno, E. and Caballero-Mora, K.S. and Capistrán, T. and Carramiñana, A. and Casanova, S. and Chaparro-Amaro, O. and Cotti, U. and Cotzomi, J. and De la Fuente, E. and Diaz Hernandez, R. and Dingus, B.L. and DuVernois, M.A. and Durocher, M. and Díaz-Vélez, J.C. and Espinoza, C. and Fan, K.L. and Fraija, N. and García-González, J.A. and Garfias, F. and González, M.M. and Goodman, J.A. and Harding, J.P. and Huang, D. and Hueyotl-Zahuantitla, F. and Iriarte, A. and Joshi, V. and Kunde, G.J. and Lee, J. and León Vargas, H. and Linnemann, J.T. and Longinotti, A.L. and Luis-Raya, G. and Lundeen, J. and Malone, K. and Martinez, O. and Martínez-Castro, J. and Matthews, J.A. and Moreno, E. and Mostafá, M. and Nayerhoda, A. and Nellen, L. and Peisker, A. and Pérez-Pérez, E.G. and Rho, C.D. and Rosa-González, D. and Salazar, H. and Salazar-Gallegos, D. and Sandoval, A. and Serna-Franco, J. and Springer, R.W. and Tibolla, O. and Tollefson, K. and Torres, I. and Torres-Escobedo, R. and Turner, R. and Ureña-Mena, F. and Villaseñor, L. and Wang, X. and Zhou, H. and de León, C.},
title = {An optimized search for dark matter in the galactic halo with HAWC},
journal = {\jcap},
}

@article{archer14,
doi = {10.1088/0004-637X/790/2/149},
url = {https://doi.org/10.1088/0004-637X/790/2/149},
year = {2014},
month = {jul},
publisher = {The American Astronomical Society},
volume = {790},
number = {2},
pages = {149},
author = {Archer, A. and Barnacka, A. and Beilicke, M. and Benbow, W. and Berger, K. and Bird, R. and Biteau, J. and Buckley, J. H. and Bugaev, V. and Byrum, K. and Cardenzana, J. V and Cerruti, M. and Chen, W. and Chen, X. and Ciupik, L. and Connolly, M. P. and Cui, W. and Dickinson, H. J. and Dumm, J. and Eisch, J. D. and Falcone, A. and Federici, S. and Feng, Q. and Finley, J. P. and Fleischhack, H. and Fortson, L. and Furniss, A. and Galante, N. and Griffin, S. and Griffiths, S. T. and Grube, J. and Gyuk, G. and Håkansson, N. and Hanna, D. and Holder, J. and Hughes, G. and Johnson, C. A. and Kaaret, P. and Kar, P. and Kertzman, M. and Khassen, Y. and Kieda, D. and Krawczynski, H. and Kumar, S. and Lang, M. J. and Maier, G. and McArthur, S. and McCann, A. and Meagher, K. and Moriarty, P. and Mukherjee, R. and Nieto, D. and O'Faoláin de Bhróithe, A. and Ong, R. A. and Otte, A. N. and Park, N. and Perkins, J. S. and Pohl, M. and Popkow, A. and Prokoph, H. and Pueschel, E. and Quinn, J. and Ragan, K. and Rajotte, J. and Reyes, L. C. and Reynolds, P. T. and Richards, G. T. and Roache, E. and Sembroski, G. H. and Shahinyan, K. and Smith, A. W. and Staszak, D. and Telezhinsky, I. and Tucci, J. V. and Tyler, J. and Varlotta, A. and Vincent, S. and Wakely, S. P. and Weinstein, A. and Welsing, R. and Wilhelm, A. and Williams, D. A. and Zajczyk, A. and Zitzer, B.},
title = {VERY-HIGH ENERGY OBSERVATIONS OF THE GALACTIC CENTER REGION BY VERITAS IN 2010–2012},
journal = {\apj}
}

@ARTICLE{berge07,
       author = {{Berge}, D. and {Funk}, S. and {Hinton}, J.},
        title = "{Background modelling in very-high-energy {\ensuremath{\gamma}}-ray astronomy}",
      journal = {\aap},
         year = 2007,
        month = may,
       volume = {466},
       number = {3},
        pages = {1219-1229},
          doi = {10.1051/0004-6361:20066674},
       url = {https://doi.org/10.1051/0004-6361:20066674}
}

@ARTICLE{bergstrom98,
title = {Observability of γ rays from dark matter neutralino annihilations in the Milky Way halo},
journal = {Astropart. Phys.},
volume = {9},
number = {2},
pages = {137-162},
year = {1998},
issn = {0927-6505},
doi = {https://doi.org/10.1016/S0927-6505(98)00015-2},
url = {https://www.sciencedirect.com/science/article/pii/S0927650598000152},
author = {Lars Bergström and Piero Ullio and James H. Buckley},
}

@PHDTHESIS{buchovecky19,
       author = {{Buchovecky}, Matthew},
        title = "{Very High Energy Emission from the Galactic Center with VERITAS}",
       school = {University of California, Los Angeles},
         year = 2019,
        month = jan,
        url = {https://www.proquest.com/docview/2284673678}
}

@ARTICLE{cirelli11,
doi = {10.1088/1475-7516/2011/03/051},
url = {https://doi.org/10.1088/1475-7516/2011/03/051},
year = {2011},
month = {mar},
publisher = {},
volume = {2011},
number = {03},
pages = {051},
author = {Marco Cirelli and Gennaro Corcella and Andi Hektor and Gert Hütsi and Mario Kadastik and Paolo Panci and Martti Raidal and Filippo Sala and Alessandro Strumia},
title = {PPPC 4 DM ID: a poor particle physicist cookbook
for dark matter indirect detection},
journal = {\jcap}
}

@INPROCEEDINGS{cogan08,
       author = {{Cogan}, P.},
        title = "{VEGAS, the VERITAS Gamma-ray Analysis Suite}",
    booktitle = {30th International Cosmic Ray Conference},
         year = 2008,
        month = jan,
        pages = {1385-1388},
archivePrefix = {arXiv},
       eprint = {0709.4233}
}

@ARTICLE{cousins08,
title = {Evaluation of three methods for calculating statistical significance when incorporating a systematic uncertainty into a test of the background-only hypothesis for a Poisson process},
journal = {Nucl. Instrum. Methods Phys. Res., Sect. A},
volume = {595},
number = {2},
pages = {480-501},
year = {2008},
issn = {0168-9002},
doi = {10.1016/j.nima.2008.07.086},
url = {https://doi.org/10.1016/j.nima.2008.07.086},
author = {Robert D. Cousins and James T. Linnemann and Jordan Tucker}
}

@ARTICLE{cowan11,
       author = {{Cowan}, Glen and {Cranmer}, Kyle and {Gross}, Eilam and {Vitells}, Ofer},
        title = "{Asymptotic formulae for likelihood-based tests of new physics}",
      journal = {Eur. Phys. J. C},
         year = 2011,
        month = feb,
       volume = {71},
          eid = {1554},
        pages = {1554},
          doi = {10.1140/epjc/s10052-011-1554-0},
          url = {https://doi.org/10.1140/epjc/s10052-011-1554-0}
}

@ARTICLE{einasto65,
       author = {{Einasto}, J.},
        title = "{On the Construction of a Composite Model for the Galaxy and on the Determination of the System of Galactic Parameters}",
      journal = {Tr. Astrofiz. Inst. Alma-Ata},
         year = 1965,
        month = jan,
       volume = {5},
        pages = {87-100},
       adsurl = {https://ui.adsabs.harvard.edu/abs/1965TrAlm...5...87E}
}

@ARTICLE{feldman98,
  title = {Unified approach to the classical statistical analysis of small signals},
  author = {Feldman, Gary J. and Cousins, Robert D.},
  journal = {Phys. Rev. D},
  volume = {57},
  issue = {7},
  pages = {3873--3889},
  numpages = {0},
  year = {1998},
  month = {Apr},
  publisher = {American Physical Society},
  doi = {10.1103/PhysRevD.57.3873},
  url = {https://link.aps.org/doi/10.1103/PhysRevD.57.3873}
}

@ARTICLE{goodenough09,
       author = {{Goodenough}, Lisa and {Hooper}, Dan},
        title = "{Possible Evidence For Dark Matter Annihilation In The Inner Milky Way From The Fermi Gamma Ray Space Telescope}",
      journal = {},
         year = 2009,
        month = oct,
archivePrefix = {arXiv},
       eprint = {0910.2998}
}

@ARTICLE{abdallah16,
  title = {Search for Dark Matter Annihilations towards the Inner Galactic Halo from 10 Years of Observations with H.E.S.S.},
  author = {Abdallah, H. and Abramowski, A. and Aharonian, F. and Ait Benkhali, F. and Akhperjanian, A. G. and Ang\"uner, E. and Arrieta, M. and Aubert, P. and Backes, M. and Balzer, A. and Barnard, M. and Becherini, Y. and Becker Tjus, J. and Berge, D. and Bernhard, S. and Bernl\"ohr, K. and Birsin, E. and Blackwell, R. and B\"ottcher, M. and Boisson, C. and Bolmont, J. and Bordas, P. and Bregeon, J. and Brun, F. and Brun, P. and Bryan, M. and Bulik, T. and Capasso, M. and Carr, J. and Casanova, S. and Chakraborty, N. and Chalme-Calvet, R. and Chaves, R. C. G. and Chen, A. and Chevalier, J. and Chr\'etien, M. and Colafrancesco, S. and Cologna, G. and Condon, B. and Conrad, J. and Couturier, C. and Cui, Y. and Davids, I. D. and Degrange, B. and Deil, C. and deWilt, P. and Djannati-Ata\"{\i}, A. and Domainko, W. and Donath, A. and Drury, L. O'C. and Dubus, G. and Dutson, K. and Dyks, J. and Dyrda, M. and Edwards, T. and Egberts, K. and Eger, P. and Ernenwein, J.-P. and Eschbach, S. and Farnier, C. and Fegan, S. and Fernandes, M. V. and Fiasson, A. and Fontaine, G. and F\"orster, A. and Funk, S. and F\"u\ss{}ling, M. and Gabici, S. and Gajdus, M. and Gallant, Y. A. and Garrigoux, T. and Giavitto, G. and Giebels, B. and Glicenstein, J. F. and Gottschall, D. and Goyal, A. and Grondin, M.-H. and Grudzi\ifmmode \acute{n}\else \'{n}\fi{}ska, M. and Hadasch, D. and Hahn, J. and Hawkes, J. and Heinzelmann, G. and Henri, G. and Hermann, G. and Hervet, O. and Hillert, A. and Hinton, J. A. and Hofmann, W. and Hoischen, C. and Holler, M. and Horns, D. and Ivascenko, A. and Jacholkowska, A. and Jamrozy, M. and Janiak, M. and Jankowsky, D. and Jankowsky, F. and Jingo, M. and Jogler, T. and Jouvin, L. and Jung-Richardt, I. and Kastendieck, M. A. and Katarzy\ifmmode \acute{n}\else \'{n}\fi{}ski, K. and Katz, U. and Kerszberg, D. and Kh\'elifi, B. and Kieffer, M. and King, J. and Klepser, S. and Klochkov, D. and Klu\ifmmode \acute{z}\else \'{z}\fi{}niak, W. and Kolitzus, D. and Komin, Nu. and Kosack, K. and Krakau, S. and Kraus, M. and Krayzel, F. and Kr\"uger, P. P. and Laffon, H. and Lamanna, G. and Lau, J. and Lees, J.-P. and Lefaucheur, J. and Lefranc, V. and Lemi\`ere, A. and Lemoine-Goumard, M. and Lenain, J.-P. and Leser, E. and Lohse, T. and Lorentz, M. and Lui, R. and Lypova, I. and Marandon, V. and Marcowith, A. and Mariaud, C. and Marx, R. and Maurin, G. and Maxted, N. and Mayer, M. and Meintjes, P. J. and Menzler, U. and Meyer, M. and Mitchell, A. M. W. and Moderski, R. and Mohamed, M. and Mor\aa{}, K. and Moulin, E. and Murach, T. and de Naurois, M. and Niederwanger, F. and Niemiec, J. and Oakes, L. and Odaka, H. and Ohm, S. and \"Ottl, S. and Ostrowski, M. and Oya, I. and Padovani, M. and Panter, M. and Parsons, R. D. and Paz Arribas, M. and Pekeur, N. W. and Pelletier, G. and Petrucci, P.-O. and Peyaud, B. and Pita, S. and Poon, H. and Prokhorov, D. and Prokoph, H. and P\"uhlhofer, G. and Punch, M. and Quirrenbach, A. and Raab, S. and Reimer, A. and Reimer, O. and Renaud, M. and de los Reyes, R. and Rieger, F. and Romoli, C. and Rosier-Lees, S. and Rowell, G. and Rudak, B. and Rulten, C. B. and Sahakian, V. and Salek, D. and Sanchez, D. A. and Santangelo, A. and Sasaki, M. and Schlickeiser, R. and Sch\"ussler, F. and Schulz, A. and Schwanke, U. and Schwemmer, S. and Seyffert, A. S. and Shafi, N. and Simoni, R. and Sol, H. and Spanier, F. and Spengler, G. and Spie\ss{}, F. and Stawarz, L. and Steenkamp, R. and Stegmann, C. and Stinzing, F. and Stycz, K. and Sushch, I. and Tavernet, J.-P. and Tavernier, T. and Taylor, A. M. and Terrier, R. and Tluczykont, M. and Trichard, C. and Tuffs, R. and van der Walt, J. and van Eldik, C. and van Soelen, B. and Vasileiadis, G. and Veh, J. and Venter, C. and Viana, A. and Vincent, P. and Vink, J. and Voisin, F. and V\"olk, H. J. and Vuillaume, T. and Wadiasingh, Z. and Wagner, S. J. and Wagner, P. and Wagner, R. M. and White, R. and Wierzcholska, A. and Willmann, P. and W\"ornlein, A. and Wouters, D. and Yang, R. and Zabalza, V. and Zaborov, D. and Zacharias, M. and Zdziarski, A. A. and Zech, A. and Zefi, F. and Ziegler, A. and \ifmmode \dot{Z}\else \.{Z}\fi{}ywucka, N.},
  collaboration = {H.E.S.S. Collaboration},
  journal = {Phys. Rev. Lett.},
  volume = {117},
  issue = {11},
  pages = {111301},
  numpages = {6},
  year = {2016},
  month = {Sep},
  publisher = {American Physical Society},
  doi = {10.1103/PhysRevLett.117.111301},
  url = {https://link.aps.org/doi/10.1103/PhysRevLett.117.111301}
}

@ARTICLE{hess18,
        title = "{Characterising the VHE diffuse emission in the central 200 parsecs of our Galaxy with H.E.S.S.}",
       author = {{H.E.S.S. Collaboration} and {Abdalla}, H. and {Abramowski}, A. and {Aharonian}, F. and {Ait Benkhali}, F. and {Akhperjanian}, A.~G. and {Andersson}, T. and {Ang{\"u}ner}, E.~O. and {Arakawa}, M. and {Arrieta}, M. and {Aubert}, P. and {Backes}, M. and {Balzer}, A. and {Barnard}, M. and {Becherini}, Y. and {Becker Tjus}, J. and {Berge}, D. and {Bernhard}, S. and {Bernl{\"o}hr}, K. and {Blackwell}, R. and {B{\"o}ttcher}, M. and {Boisson}, C. and {Bolmont}, J. and {Bonnefoy}, S. and {Bordas}, P. and {Bregeon}, J. and {Brun}, F. and {Brun}, P. and {Bryan}, M. and {B{\"u}chele}, M. and {Bulik}, T. and {Capasso}, M. and {Carr}, J. and {Casanova}, S. and {Cerruti}, M. and {Chakraborty}, N. and {Chaves}, R.~C.~G. and {Chen}, A. and {Chevalier}, J. and {Coffaro}, M. and {Colafrancesco}, S. and {Cologna}, G. and {Condon}, B. and {Conrad}, J. and {Cui}, Y. and {Davids}, I.~D. and {Decock}, J. and {Degrange}, B. and {Deil}, C. and {Devin}, J. and {deWilt}, P. and {Dirson}, L. and {Djannati-Ata{\"\i}}, A. and {Domainko}, W. and {Donath}, A. and {Drury}, L.~O. 'C. and {Dutson}, K. and {Dyks}, J. and {Edwards}, T. and {Egberts}, K. and {Eger}, P. and {Ernenwein}, J. -P. and {Eschbach}, S. and {Farnier}, C. and {Fegan}, S. and {Fernandes}, M.~V. and {Fiasson}, A. and {Fontaine}, G. and {F{\"o}rster}, A. and {Funk}, S. and {F{\"u}{\ss}ling}, M. and {Gabici}, S. and {Gallant}, Y.~A. and {Garrigoux}, T. and {Giavitto}, G. and {Giebels}, B. and {Glicenstein}, J.~F. and {Gottschall}, D. and {Goyal}, A. and {Grondin}, M. -H. and {Hahn}, J. and {Haupt}, M. and {Hawkes}, J. and {Heinzelmann}, G. and {Henri}, G. and {Hermann}, G. and {Hinton}, J.~A. and {Hofmann}, W. and {Hoischen}, C. and {Holch}, T.~L. and {Holler}, M. and {Horns}, D. and {Ivascenko}, A. and {Iwasaki}, H. and {Jacholkowska}, A. and {Jamrozy}, M. and {Janiak}, M. and {Jankowsky}, D. and {Jankowsky}, F. and {Jingo}, M. and {Jogler}, T. and {Jouvin}, L. and {Jung-Richardt}, I. and {Kastendieck}, M.~A. and {Katarzy{\'n}ski}, K. and {Katsuragawa}, M. and {Katz}, U. and {Kerszberg}, D. and {Khangulyan}, D. and {Kh{\'e}lifi}, B. and {King}, J. and {Klepser}, S. and {Klochkov}, D. and {Klu{\'z}niak}, W. and {Kolitzus}, D. and {Komin}, Nu. and {Kosack}, K. and {Krakau}, S. and {Kraus}, M. and {Kr{\"u}ger}, P.~P. and {Laffon}, H. and {Lamanna}, G. and {Lau}, J. and {Lees}, J. -P. and {Lefaucheur}, J. and {Lefranc}, V. and {Lemi{\`e}re}, A. and {Lemoine-Goumard}, M. and {Lenain}, J. -P. and {Leser}, E. and {Lohse}, T. and {Lorentz}, M. and {Liu}, R. and {L{\'o}pez-Coto}, R. and {Lypova}, I. and {Marandon}, V. and {Marcowith}, A. and {Mariaud}, C. and {Marx}, R. and {Maurin}, G. and {Maxted}, N. and {Mayer}, M. and {Meintjes}, P.~J. and {Meyer}, M. and {Mitchell}, A.~M.~W. and {Moderski}, R. and {Mohamed}, M. and {Mohrmann}, L. and {Mor{\r{a}}}, K. and {Moulin}, E. and {Murach}, T. and {Nakashima}, S. and {de Naurois}, M. and {Niederwanger}, F. and {Niemiec}, J. and {Oakes}, L. and {O'Brien}, P. and {Odaka}, H. and {Ohm}, S. and {Ostrowski}, M. and {Oya}, I. and {Padovani}, M. and {Panter}, M. and {Parsons}, R.~D. and {Pekeur}, N.~W. and {Pelletier}, G. and {Perennes}, C. and {Petrucci}, P. -O. and {Peyaud}, B. and {Piel}, Q. and {Pita}, S. and {Poon}, H. and {Prokhorov}, D. and {Prokoph}, H. and {P{\"u}hlhofer}, G. and {Punch}, M. and {Quirrenbach}, A. and {Raab}, S. and {Rauth}, R. and {Reimer}, A. and {Reimer}, O. and {Renaud}, M. and {de los Reyes}, R. and {Richter}, S. and {Rieger}, F. and {Romoli}, C. and {Rowell}, G. and {Rudak}, B. and {Rulten}, C.~B. and {Sahakian}, V. and {Saito}, S. and {Salek}, D. and {Sanchez}, D.~A. and {Santangelo}, A. and {Sasaki}, M. and {Schlickeiser}, R. and {Sch{\"u}ssler}, F. and {Schulz}, A. and {Schwanke}, U. and {Schwemmer}, S. and {Seglar-Arroyo}, M. and {Settimo}, M. and {Seyffert}, A.~S. and {Shafi}, N. and {Shilon}, I. and {Simoni}, R. and {Sol}, H. and {Spanier}, F. and {Spengler}, G. and {Spies}, F. and {Stawarz}, {\L}. and {Steenkamp}, R. and {Stegmann}, C. and {Stycz}, K. and {Sushch}, I. and {Takahashi}, T. and {Tavernet}, J. -P. and {Tavernier}, T. and {Taylor}, A.~M. and {Terrier}, R. and {Tibaldo}, L. and {Tiziani}, D. and {Tluczykont}, M. and {Trichard}, C. and {Tsuji}, N. and {Tuffs}, R. and {Uchiyama}, Y. and {van der Walt}, D.~J. and {van Eldik}, C. and {van Rensburg}, C. and {van Soelen}, B. and {Vasileiadis}, G. and {Veh}, J. and {Venter}, C. and {Viana}, A. and {Vincent}, P. and {Vink}, J. and {Voisin}, F. and {V{\"o}lk}, H.~J. and {Vuillaume}, T. and {Wadiasingh}, Z. and {Wagner}, S.~J. and {Wagner}, P. and {Wagner}, R.~M. and {White}, R. and {Wierzcholska}, A. and {Willmann}, P. and {W{\"o}rnlein}, A. and {Wouters}, D. and {Yang}, R. and {Zaborov}, D. and {Zacharias}, M. and {Zanin}, R. and {Zdziarski}, A.~A. and {Zech}, A. and {Zefi}, F. and {Ziegler}, A. and {{\.Z}ywucka}, N.},
      journal = {\aap},
         year = 2018,
        month = apr,
       volume = {612},
          eid = {A9},
        pages = {A9},
          doi = {10.1051/0004-6361/201730824},
       url = {https://doi.org/10.1051/0004-6361/201730824}
}

@ARTICLE{hillas96,
       author = {{Hillas}, A.~M.},
        title = "{Differences between Gamma-Ray and Hadronic Showers}",
      journal = {\ssr},
         year = 1996,
        month = jan,
       volume = {75},
       number = {1-2},
        pages = {17-30},
          doi = {10.1007/BF00195021},
       adsurl = {https://ui.adsabs.harvard.edu/abs/1996SSRv...75...17H}
}

@INPROCEEDINGS{hona21,
       author = {Hona, Binita},
        title = "{Matched Runs Method to Study Extended Regions of Gamma-ray Emission}",
    booktitle = {37th International Cosmic Ray Conference},
         year = {2021},
        pages = {729},
       volume = {395},
          doi = {10.22323/1.395.0729},
       eprint = {2108.07663}
}

@PHDTHESIS{mayer14,
  author = {{Mayer}, Michael},
  title = {Pulsar Wind Nebulae at High Energies: A Diverse Population and Exceptional Twins},
  school = {Potsdam University},
  year = 2014,
  url = {https://bib-pubdb1.desy.de/record/393653}
}

\end{document}